\RequirePackage{fix-cm}

\documentclass[pre,reprint,twocolumn,showpacs,floatfix]{revtex4-1}

\usepackage{textcomp}

\usepackage{color}
\usepackage{amsmath}
\usepackage{amssymb}
\usepackage{graphicx}
\usepackage{nicefrac}

\newcommand{\figWidth}{0.95\columnwidth}

\newcommand{\eqRef}[1]{Eq.~(\ref{#1})}
\newcommand{\figRef}[1]{Fig.~\ref{#1}}
\newcommand{\secRef}[1]{Sec.~\ref{#1}}
\newcommand{\tabRef}[1]{Tab.~\ref{#1}}

\begin{document}

\title{Event-driven Molecular Dynamics of Soft Particles}
\author{Patric M\"uller and Thorsten P\"oschel}

\affiliation{Institute for Multiscale Simulation, Universit\"at Erlangen-N\"urnberg, N\"agelsbachstra{\ss}e 49b,
91052 Erlangen, Germany}

\date{\today}

\begin{abstract}

The dynamics of dissipative soft-sphere gases obeys Newton's equation of motion which are commonly solved numerically by (force-based) Molecular Dynamics schemes. With the assumption of instantaneous, pairwise collisions, the simulation can be accelerated considerably using event-driven Molecular Dynamics, where the coefficient of restitution is derived from the interaction force between particles. Recently it was shown, however, that this approach may fail dramatically, that is, the obtained trajectories deviate significantly from the ones predicted by Newton's equations. In this paper, we generalize the concept of the coefficient of restitution and derive a numerical scheme which, in the case of dilute systems and frictionless interaction, allows us to perform highly efficient event-driven Molecular Dynamics simulations even for non-instantaneous collisions. We show that the particle trajectories predicted by the new scheme agree perfectly with the corresponding (force-based) Molecular Dynamics, except for a short transient period whose duration corresponds to the duration of the contact. Thus, the new algorithm solves Newton's equations of motion like force-based MD while preserving the advantages of event-driven simulations.
\end{abstract}
   
\maketitle
\section{Introduction}
\label{sec:intro}
Modelling granular systems of frictionless spheres branches into two fundamental different approaches: Hard- and soft-sphere models. The dynamics of soft spheres are governed by the pairwise interaction forces between contacting particles as a function of the relative particle positions and velocities as well as material parameters, $\vec{F}_{ij}=\vec{F}_{ij}\left(\vec{r}_i, \vec{r}_j, \dot{\vec{r}}_i, \dot{\vec{r}}_j\right)$. The dynamics of a many-particle system is then obtained by numerically solving Newton's equation of motion for all degrees of freedom, which was termed Molecular Dynamics (MD), e.g. \cite{allen1987}. The first MD simulations of granular systems (in the engineering literature also called {\em Discrete Element Method -- DEM}) range back to pioneering work by Cundall, Walton, Haff and others, e.g. \cite{cundall1979,gallas1992,haff1986,walton1982}. An overview of the force models specific for granular particles can be found in \cite{kruggelEmden2007,stevens2005,schaefer1996}.

In contrast to soft-sphere models, in hard-sphere models the collisions are assumed to occur \emph{instantaneously} which allows to consider the dynamics of hard sphere systems as a sequence of independent binary collisions. Except for collisions where the velocities change instantaneously, the particles follow ballistic trajectories, possibly under the influence of external fields like gravity.
The hard sphere model is the foundation of both, Kinetic Theory of granular matter based on the Boltzmann equation e.g. \cite{goldhirsch2003,GranularGases,PROCEEDINGS}, and event-driven Molecular Dynamics (eMD) of granular matter, e.g. \cite{lubachevsky1991,rapaport1980,poeschel2005}. 

The collision of two hard spheres of velocities  $\dot{\vec{r}}_i$ and $\dot{\vec{r}}_j$ implies an instantaneous exchange of momentum:
\begin{equation}
\label{eq:epsNCorrectDef}
 \left(\dot{\vec{r}}_i^{\,\prime}-\dot{\vec{r}}_j^{\,\prime}\right)\cdot\vec{e}^{\:\prime}_{r}
=-\varepsilon_n\left(\dot{\vec{r}}^{\,0}_i-\dot{\vec{r}}_j^{\,0}\right)
\cdot\vec{e}_{r}^{\,0}\,
\end{equation}
with the time dependent inter center unit vector $\vec{e}_{r}\equiv \left(\vec{r}_i-\vec{r}_j\right)/\left|\vec{r}_i-\vec{r}_j\right|$ and the coefficient of normal restitution $\varepsilon_n$. Upper index $0$ denotes values just before the collision, primed values denote post-collisional values. Unlike the velocities, the particles' positions remain unchanged because of the instantaneous character of the collision, therefore,
\begin{equation}
\label{eq:fixedPos}
\vec{e}^{\,\prime}_{r}\equiv\vec{e}_{r}^{\,0}\:
\end{equation}
and \eqRef{eq:epsNCorrectDef} reduces to
\begin{equation}
\label{eq:epsNHSDef}
 \left(\dot{\vec{r}}_i^{\,\prime}-\dot{\vec{r}}_j^{\,\prime}\right)\cdot\vec{e}^{\,0}_{r}
=-\varepsilon_n\left(\dot{\vec{r}}^{\,0}_i-\dot{\vec{r}}_j^{\,0}\right)
\cdot\vec{e}_{r}^{\,0}\:.
\end{equation}
Equation~\eqref{eq:epsNHSDef} relating the pre- and post-collisional velocities is the governing equation of eMD.  
Given a certain granular system may be described by the hard-sphere model, eMD allows for a vast increase of numerical efficiency as compared with corresponding MD simulations. For a very efficient implementation of eMD see \cite{Bannerman}.

Despite of eMD's great numerical performance, the hard-sphere model is a simplification of physical reality: instantaneous changes of velocity imply infinite delta-shaped forces while forces between colliding physical objects are always finite which implies finite contact duration. Therefore, the applicability of the hard-sphere model for eMD simulations of granular systems must be checked. One obvious precondition for eMD is low enough particle number density such that the frequency of three-particle contacts can be neglected as compared to the frequency of pair collisions. Obviously, this is not given for slow flows with long-lasting contacts. 

A natural way to check the validity of the hard-sphere approximation in the dilute limit is the following: The coefficient of normal restitution as a function of material parameters and relative impact velocity may be obtained from analytically integrating Newton's equation of motion for the central collision of an isolated pair of particles using the known interaction force which is also a function of material properties and impact velocity, e.g. \cite{schwager2007,schwager2008,ramirez1999,Schwager:1998}. Performing now MD simulations using the interaction force and eMD simulations using the corresponding expression for the coefficient of restitution, one may expect identical trajectories. However, recently it was found that these trajectories may deviate significantly for a vast range of materials, collision geometries and impact velocities \cite{mueller2011,negCONR}, in particular for oblique impacts which concerns the majority of impact geometries \cite{mueller2011} in a Molecular Chaos situation. Consequently, even for dilute systems, the hard-sphere approximation may fail dramatically. This effect may be attributed to the finite duration of collisions in physical systems which does not allow for the assumption Eq. \eqref{eq:fixedPos}.

Consequently, on one hand we have the stunning efficiency of eMD based on the hard-sphere model. On the other hand there is the universality and physical correctness of the soft-sphere model leading to MD. Combining the advantages of both approaches is a highly desired aim. Concerning simulation techniques an attempt is to discretize the (smooth) interaction potentials. As this idea was originally developed for liquids \cite{chapela1984,chapela1989} recently it was also applied to granular systems \cite{goyal2010,deLaPena2007,vanZon2008}. On the theoretical side there are perturbation theories, extending hard sphere models \cite{barker1967,weeks1970,chapela2010}.  

In this work we derive an algorithm for the event-driven simulation of smooth spheres which does not rely on Eq. \eqref{eq:fixedPos}. By extending the concept of the coefficient of restitution, we map the correct Newtonian dynamics of soft spheres to instantaneous events. We show that for dilute systems of frictionless particles the presented method allows for a correct computation of the trajectories (as MD) while preserving the efficiency of event-driven simulations. 

This simulation method applies to a wide range of particle interaction forces. Here we demonstrate it for the case of two important examples: The linear dashpot model and viscoelastic spheres. Unlike the original eMD method, we show that in these cases the trajectories obtained by eMD agree perfectly with the MD results.

% Within an eMD simulation we measure the coefficient of self diffusion of a granular gas. Once using the hard sphere collision rule \eqRef{eq:epsNHSDef} and once using our collision mapping. If applicable, it exactly reassembles the dynamics of soft spheres. The difference between both measurements reveals sever deficiencies of the hard sphere model and calls for reconsideration of Kinetic Theory, as well as event-driven simulation of granular matter.      

\section{Collision of spheres}
\label{sec:smoothSpheres}
Consider two colliding spheres of masses $m_i$ and $m_j$ located at $\vec{r}_i(t)$ and $\vec{r}_j(t)$ and traveling with velocities $\dot{\vec{r}}_i(t)$ and $\dot{\vec{r}}_j(t)$.
With the interaction force $\vec{F}$, their motion is described by 
\begin{equation}
\label{eq:newton}
m_{\text{eff}}\,\ddot{\vec{r}}=\vec{F}\,,~~~~~
M\ddot{\vec{R}}=\vec{0}
\end{equation}
where 
\begin{equation}
\label{eq:COMAndRelCoordDef}
 \vec{R}\equiv \frac{m_i\vec{r}_i+m_j\vec{r}_j}{m_i+m_j}\,,~~~\vec{r}=\vec{r}_i-\vec{r}_j\,,~~~m_{\text{eff}}=\frac{m_im_j}{m_i+m_j}
\end{equation}
are the center of mass coordinate, the relative coordinate and the effective mass, respectively. The center of mass moves due to external forces such as gravity and separates from the relative motion which in turn contains the entire collision dynamics. 

For frictionless particles, the interaction force acts in the direction of the inter-center unit vector, $\vec{F}=F_n \vec{e}_r$.
%, that is, there is no tangential force and, thus, the particles' rotation is not affected by the collision. 
During the collision the (orbital) angular momentum is conserved which allows for the definition of the constant unit vector $\vec{e}_L$:
\begin{equation}
\label{eq:angMomDef}
 \vec{L}= m_{\text{eff}}\:\vec{r}\times\dot{\vec{r}}\equiv L\vec{e}_L\,.
\end{equation}
Thus, with the coordinate system $\Sigma$ spanned by
\begin{equation}
\label{eq:polarDef}
\vec{e}_{x}\equiv\vec{e}_r^{\,0}\,,~~~~
\vec{e}_{z}\equiv\vec{e}_L\,,~~~~
\vec{e}_{y}\equiv\vec{e}_z\times\vec{e}_x\,,
\end{equation}
and with its origin in the center of mass $\vec{R}$, the collision takes place in the $\vec{e}_{x}$-$\vec{e}_{y}$--plane \footnote{For central collisions we have $\vec{L}=\vec{0}$. In this case $\vec{e}_z$ may be any unit vector perpendicular to $\vec{e}_x$, ($\vec{e}_x\cdot\vec{e}_z=0$).}. 
\begin{figure}[h!]
 \includegraphics[width=\figWidth]{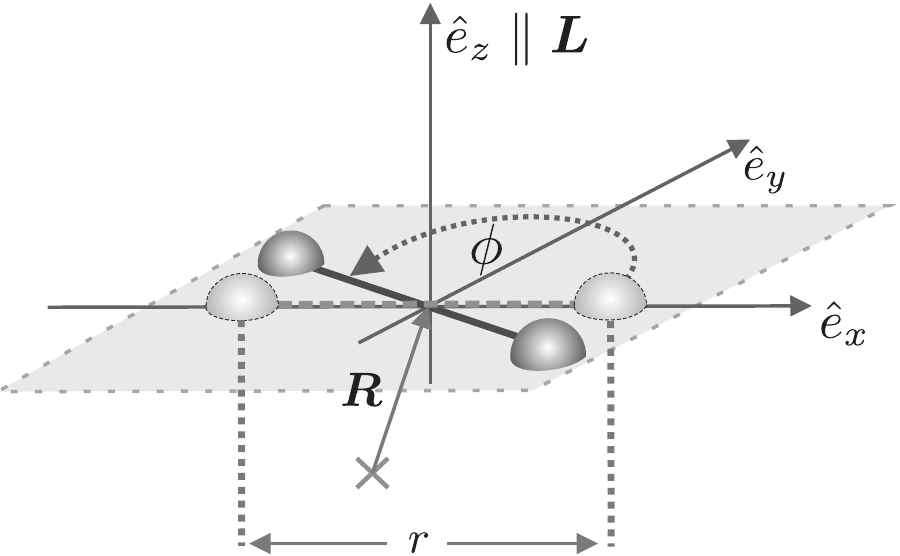}
 \caption{Illustration of the used polar coordinates (see text)}
 \label{fig:polarDef}
\end{figure}
In the collision plane we formulate the equation of motion in polar coordinates
$\left\{r,\varphi\right\}$ (see \figRef{fig:polarDef}):
\begin{equation}
\label{eq:newtonPolar}
 m_{\text{eff}}\,r^2\dot{\varphi}=L\,,~~~~
 m_{\text{eff}}\,\ddot{r}=F_c+F_n=m_{\text{eff}}\,r\dot{\varphi}^2+F_n\,,
\end{equation}
with the centrifugal force $F_c$. Together with the inital conditions 
\begin{equation}
\label{eq:initCond}
 r(0)=r^0\,,~~~~
 \dot{r}(0)=\dot{r}^0\,,~~~~
 \varphi(0)=0\,,
\end{equation}
\eqRef{eq:newtonPolar} fully describes the collision dynamics for an arbitrary
normal force $F_n$. The collision terminates at time $t=\tau$ where \cite{schwager2007,schwager2008}
\begin{equation}
\label{eq:tauDef}
 \dot{r}(\tau)>0\text{~~~and~~~}F_n=0.
\end{equation}

Measuring time in units of $T$, length in units of $X$ and angles in units of $\varPhi$, and using the dimensionless quantities 
\begin{equation}
\label{eq:scaleDef}
 \tilde{r}=\frac{r}{X},\quad\tilde{t}=\frac{t}{T}\quad\text{and }\tilde{\varphi}=\frac{\varphi}{\varPhi}
\end{equation}
we obtain the scaled form of the equation of motion, \eqRef{eq:newtonPolar}:
\begin{eqnarray}
\label{eq:newtonPolarScaled}
 \frac{\mathrm d \tilde{\varphi}}{\mathrm d
\tilde{t}}&=&\frac{c_\varphi}{\tilde{r}^2}\nonumber\\
 \frac{\mathrm d^2\tilde{r}}{\mathrm d\tilde{t}^2}&=&
 \tilde{r}\left(\frac{\mathrm d \tilde{\varphi}}{\mathrm d \tilde{t}}\right)^2\varPhi^2+
 \frac{F_n}{m_{\text{eff}}}\frac{T^2}{X},
\end{eqnarray}
where $X$ and $T$ are length and time scales typical for the given normal force $F_n$, $\varPhi$ is an arbitrary scale for measuring angles and $c_\varphi$ reads
\begin{equation}
 \label{eq:cPhiDef}
 c_\varphi=\frac{T}{\varPhi X^2}\frac{L}{m_{\text{eff}}}.
\end{equation}
The corresponding dimensionless initial conditions read
\begin{equation}
\label{eq:initCondScaled}
 \tilde{\varphi}(0)=0,\quad\tilde{r}(0)=\frac{r(0)}{X}\quad\text{and}\quad\frac{\mathrm d \tilde{r}}{\mathrm d \tilde{t}}(0)=\dot{r}(0)\frac{T}{X}.
\end{equation}
According to \eqRef{eq:tauDef} the scaled contact duration reads $\tilde{\tau}\equiv\tau/T$.

\section{Coefficient of restitution versus Matrix of Restitution}
\label{sec:collMapping}

Solving the scaled equation of motion \eqRef{eq:newtonPolarScaled} for the initial conditions \eqRef{eq:initCondScaled} in the time interval $0\le \tilde{t}\le \tilde{\tau}$, that is, from the beginning of the collision at time $\tilde{t}=0$ until its end at $\tilde{t}=\tilde{\tau}$ (see \eqRef{eq:tauDef}) we obtain the post-collisional values $\tilde{\varphi}(\tilde{\tau})$, $\dot{\tilde{\varphi}}(\tilde{\tau})$, $\tilde{r}(\tilde{\tau})$ and $\dot{\tilde{r}}(\tilde{\tau})$, which determine the state of the system at the end of the collision. 

Note that for the special case of central collisions with vanishing angular momentum, the state would be fully described by $\tilde{r}(\tilde{\tau})$ and $\dot{\tilde{r}}(\tilde{\tau})$ as the other values vanish or are invariant. Together with the hard-spere assumption, Eq. \eqref{eq:fixedPos}, we are left with only $\dot{\tilde{r}}(\tilde{\tau})$ which allow us to caracterize the collision by a single number, the coefficient of restitution,
\begin{equation}
  \label{eq:1}
  \varepsilon_n = -\frac{\dot{\tilde{r}}(\tilde{\tau})}{\dot{\tilde{r}}(0)}\,.
\end{equation}
Therefore, $\varepsilon_n \in [0,1]$ for central collisions.

Equation \eqref{eq:1} provides the link between the hard-sphere and the soft-sphere models and, correspondingly, between MD and eMD since it relates the coefficient of restitution with the specific interaction force. The analytical solution of Eq. \eqref{eq:1} is frequently non-trivial, even for rather simple forces as the viscoelastic Hertz force \cite{schwager2008,Schwager:1998}. As a result from the solution of Eq.~\eqref{eq:1} we obtain the coefficient of normal restitution as a function of the force's material specifics, particle sizes and impact rate. 

Obviously, only for the special case of central collisions of vanishing duration, $\varepsilon_n$ is sufficient to characterize collisions since otherwise $\tilde{\varphi}(\tilde{\tau})$, $\dot{\tilde{\varphi}}(\tilde{\tau})$ and  $\tilde{r}(\tilde{\tau})$ do not vanish. It was shown for ordinary material and impact parameters, that the mentioned post-collisional quantities are not negligible \cite{mueller2011}. If one overrides this fact, the coefficient of restitution {\em must} depend on the impact parameter $d$ (see Fig. \ref{fig:excCollSetup} below). Depending on $d$ it can adopt even negative values \cite{negCONR}. Therefore, we believe that knowing the coefficient of restitution, $\varepsilon_n$, is not sufficient to perform particle simulations. 

Following the previous arguments, besides the ordinary coefficient of normal restitution, we define further coefficients which together characterize the collision completely. These are
\begin{equation}
\label{eq:epsCompDefA}
 \varepsilon_r\equiv \frac{\tilde{r}(\tilde{\tau})}{\tilde{r}(0)}\,,
\end{equation}
which stands for distance of the colliders at the end of the collision.  Na\"{\i}vely one could believe $\varepsilon_r=1$ since the particles lose contact when $\left|\vec{r}_i-\vec{r}_j\right| = R_i+R_j$. However, as shown in \cite{schwager2008,schwager2007}, the latter condition is not correct and leads to erroneous attractive forces even if the interaction force between the particles was assumed purely repulsive. In fact, $\varepsilon_r \lesssim 1$.

The next coefficient, 
\begin{equation}
\label{eq:epsCompDefB}
 \varepsilon_{\varphi}\equiv\tilde{\varphi}(\tilde{\tau})
\end{equation}
represents the rotation of the normal vector $\vec{e}_r$ during the collision, measured in units of $\varPhi$. It is defined by $\vec{e}_r^{\,0}\cdot\vec{e}_r^{\,\prime}=\cos(\varepsilon_\varphi\varPhi)$.  

The change of the corresponding rotation velocity is described by a further coefficient, 
\begin{equation}
\label{eq:epsCompDefC}
 \varepsilon_{\dot{\varphi}}\equiv\frac{\dot{\tilde{\varphi}}(\tilde{\tau})}{\dot{\tilde{\varphi}}(0)}\,.
\end{equation}
Using the conservation of angular momentum, $L=m_{\text{eff}\,}r^2(t)\dot{\varphi}(t)$, we see that this coefficient is redundant and may be expressed through $\varepsilon_r$:
\begin{equation}
\label{eq:epsPhiDotDef}
 \varepsilon_{\dot{\varphi}}=\left(\frac{\tilde{r}(0)}{\tilde{r}(\tilde{\tau})}\right)^2=\left(\frac{1}{\varepsilon_r}\right)^2\,. 
\end{equation}
The propagation of time is accounted for by 
\begin{equation}
\label{eq:epsCompDefD}
 \varepsilon_{t}\equiv\tilde{\tau}
\end{equation}
which holds the scaled contact time. It is obviously needed since {\em time} is also a variable which changes during a mechanical contact. Its meaning becomes clear if one looks to the center of mass coordinate $\vec{R}$ which is not affected by the collision due to momentum conservation. To determine its post-collisional value, one needs to know the time when the collision terminates.

Finally we need 
\begin{equation}
\label{eq:epsCompDefE}
 \varepsilon_{\dot{r}}\equiv\frac{\dot{\tilde{r}}(\tilde{\tau})}{\dot{\tilde{r}}(0)}
\end{equation}
which is (up to the sign) the ordinary coefficient of normal restitution including the influence of centrifugal forces occurring for non central collisions, $-\varepsilon_{\dot{r}}=\varepsilon_n$.

Following the arguments of the previous section, the state of the colliding particles is completely determined by 
$r(t)$, $\dot{r}(t)$, $\varphi(t)$, $\dot{\varphi}(t)$ and $t$. If we define
\begin{equation}
\vec{\chi}(0)\equiv\left(
\begin{array}{c}
r^{\:0}\\
\dot{r}^{\:0}\\
\varPhi\\
\dot{\varphi}^{\:0}\\
T
\end{array}
\right) \,,
\end{equation}
equations \eqref{eq:epsCompDefA}-\eqref{eq:epsCompDefE}  establish then a complete set of equations to compute the post-collisional state, $\vec{\chi}(\tau)$, from the pre-collisional one, $\vec{\chi}(0)$

% \begin{equation}
% \label{eq:epsCompDef}
% \begin{split}
%   \varepsilon_r&\equiv \frac{\tilde{r}(\tilde{\tau})}{\tilde{r}(0)}\\
%   \varepsilon_{\dot{r}}&\equiv\frac{\dot{\tilde{r}}(\tilde{\tau})}{\dot{\tilde{r}}(0)}\\
%   \varepsilon_{\varphi}&\equiv\tilde{\varphi}(\tilde{\tau})\\
%   \varepsilon_{\dot{\varphi}}&\equiv\frac{\dot{\tilde{\varphi}}(\tilde{\tau})}{\dot{\tilde{\varphi}}(0)}\\
%   \varepsilon_{t}&\equiv\tilde{\tau}\,.
% \end{split}
% \end{equation}

We arrange the coefficients given in Eqs.~\eqref{eq:epsCompDefA}-\eqref{eq:epsCompDefE} in form of the {\em matrix of restitution} 
\begin{equation}
\label{eq:colMat}
\tilde{\varepsilon}=\left(
\begin{array}{ccccc}
\varepsilon_r&0&0&0&0\\
0&\varepsilon_{\dot{r}}&0&0&0\\
0&0&\varepsilon_\varphi&0&0\\
0&0&0&\nicefrac{1}{\varepsilon_r^2}&0\\
0&0&0&0&\varepsilon_t
\end{array}
\right),
\end{equation}
such that the collision dynamics is described by the propagator
\begin{equation}
\label{eq:cRule}
\vec{\chi}(\tau)=\tilde{\varepsilon}\,\vec{\chi}(0)\,,
\end{equation}
which has exactly the same functional form as a the traditional propagator rule, Eq. \eqref{eq:1}

Similar to Eq. \eqref{eq:1} which is the basic equation of eMD under the simplifying assumption, Eq. \eqref{eq:fixedPos}, of instantaneous collisions, 
Eq. \ref{eq:cRule} will be the basic equation of our generalized eMD, which does not rely in instantaneous collisions.

\section{Improved Collision Rule}
\label{sec:cRule}

In traditional eMD simulations the particles move along straight lines or ballistic trajectories under the influence of constant external fields like gravity, interrupted by instantaneous events (collisions) where their velocities are adjusted according to the collision law. That is, the collision law does not change the positions of the particles.

The new propagator, Eq. \eqref{eq:cRule}, requires that the corresponding collision law changes both the velocities and also the positions of the particles. The change of the position in a collision will cause some problems in simulations, namely, it may happen that the designated positions are occupied by other particles. This problem will be addressed in Section \ref{sec:impr-event-driv}. 
%\todo{In the present section, we describe the update of the particles' velocities and positions, provided the new positions are not occupied. }\patric{Ist diese todo hinfaellig?}

In the present section, we detail the update of the particles' velocities and positions, provided the new positions are not occupied. We describe how to apply the matrix of restitution $\tilde{\varepsilon}$ to obtain the post-collisional coordinates $\vec{r}^{\,\prime}_1$, $\vec{r}^{\,\prime}_2$, $\vec{v}^{\,\prime}_1$, $\vec{v}^{\,\prime}_2$ from the pre-collisional coordinates $\vec{r}_1^{\,0}$, $\vec{r}_2^{\,0}$, $\vec{v}^{\,0}_1$, $\vec{v}^{\,0}_2$ for a given set of material parameters and particle masses. For convenience, we use two (fixed) reference frames: The laboratory system $\Sigma^\text{L}$ (spanned by $\vec{e}_x^{\,\text{L}}$, $\vec{e}_y^{\,\text{L}}$, $\vec{e}_z^{\,\text{L}}$) and $\Sigma$ as defined in \secRef{sec:smoothSpheres}, \eqRef{eq:polarDef}. $\hat{X}$  indicates, that the vector $X$ is expressed in the reference frame $\Sigma$. Vectors without a hat are expressed in $\Sigma^\text{L}$, respectively. 

\subsection{Position Update}
\label{subSec:cRulePosUpdate}
%To update the particle positions, we need to switch to the reference frame $\Sigma$. 
The base vectors of the laboratory frame $\Sigma^\text{L}$ expressed in $\Sigma$ read 
\begin{equation}
\hat{\vec{e}}_i^{\,\text{L}}=
\left(
  \begin{array}{c}
\vec{e}_i^{\,\text{L}}\cdot\vec{e}_x\\
\vec{e}_i^{\,\text{L}}\cdot\vec{e}_y\\
\vec{e}_i^{\,\text{L}}\cdot\vec{e}_z  
  \end{array}
\right)\:.
\end{equation}
The direction of the relative coordinate $\vec{e}_r^{\,\prime}$ after the collision reads
\begin{equation}
 \hat{\vec{e}}_r^{\,\prime}=\left(
   \begin{array}{c}
     \cos(\varepsilon_\varphi\varPhi)\\
     \sin(\varepsilon_\varphi\varPhi)\\
     0
   \end{array}
\right),
\end{equation}
expressed in the reference frame $\Sigma$. The corresponding vector expressed in $\Sigma^\text{L}$ reads
\begin{equation}
\vec{e}_r^{\:\prime}=
\left(
  \begin{array}{c}
    \hat{\vec{e}}_r^{\,\prime}\cdot\hat{\vec{e}}_x^{\,\text{L}}\\
    \hat{\vec{e}}_r^{\,\prime}\cdot\hat{\vec{e}}_y^{\,\text{L}}\\
    \hat{\vec{e}}_r^{\,\prime}\cdot\hat{\vec{e}}_z^{\,\text{L}}    
  \end{array}
\right)
\end{equation}
The distance $r^{\prime}$ between the two spheres after the collision is given by
\begin{equation}
r^{\prime}=r^0\varepsilon_r,
\end{equation}
where $r^0$ is its precollisional value. With this, the vector pointing from the origin of $\Sigma$ to particle 1 after the collision reads
\begin{equation}
\Delta\vec{r}_1^{\,\prime}=-\frac{m_2}{m_1+m_2}r^{\prime}\:\vec{e}_r^{\,\prime},
\end{equation}
expressed in the laboratory frame $\Sigma^\text{L}$. The corresponding vector pointing to particle 2 reads
\begin{equation}
\Delta\vec{r}_2^{\,\prime}=\frac{m_1}{m_1+m_2}r^{\prime}\:\vec{e}_r^{\,\prime}.
\end{equation}
The center of mass coordinate after the collision reads
\begin{equation}
\vec{R}^{\prime}=\vec{R}^0+\dot{\vec{R}}^0\varepsilon_t T
\end{equation}
expressed in the laboratory frame.

With this, the postcollisional particle positions expressed in the laboratory frame read
\begin{equation}
\label{eq:post-positions}
\vec{r}_i^{\,\prime}=\vec{R}^{\prime}+\Delta\vec{r}_i^{\,\prime}
\end{equation}

\subsection{Velocity Update}
%Updating the velocities is done just like the position update described in \ref{subSec:cRulePosUpdate}.: 
The angular velocity  at the instant of collision is given by
\begin{equation}
 \dot{\varphi}^0=\frac{L}{m_{\text{eff}}\:{\left(r^0\right)}^2}.
\end{equation}
The corresponding postcollisional value reads
\begin{equation}
\dot{\varphi}^\prime=\frac{\dot{\varphi}(0)}{\varepsilon_r^2}.
\end{equation}
The derivative of the unit vector of the postcollisional relative coordinate reads
\begin{equation}
\hat{\vec{e}}_{\dot{r}}^{\,\prime}=
\left(
  \begin{array}{c}
-\dot{\varphi}^\prime\sin(\varepsilon_\varphi\varPhi)\\
\dot{\varphi}^\prime\cos(\varepsilon_\varphi\varPhi)\\
0    
  \end{array}
\right)
\end{equation}
in the reference frame $\Sigma$. The corresponding vector expressed in the laboratory frame $\Sigma^\text{L}$ reads
\begin{equation}
\vec{e}_{\dot{r}}^{\,\prime}=
\left(
  \begin{array}{cc}
    \hat{\vec{e}}_{\dot{r}}^{\,\prime}\cdot\hat{\vec{e}}_x^{\,\text{L}}\\
    \hat{\vec{e}}_{\dot{r}}^{\,\prime}\cdot\hat{\vec{e}}_y^{\,\text{L}}\\
    \hat{\vec{e}}_{\dot{r}}^{\,\prime}\cdot\hat{\vec{e}}_z^{\,\text{L}}
\end{array}
\right)\:.
\end{equation}
The normal component $\dot{r}^{\prime}$ of the relative velocity between the two spheres after the collision is given by
\begin{equation}
\dot{r}^{\prime}=\dot{r}^0\varepsilon_{\dot{r}},
\end{equation}
where $\dot{r}^0$ is its pre-collisional value. With this, the post-collisional velocity of 
%particle 1 
the particles measured from the origin of $\Sigma$, expressed in the laboratory frame $\Sigma^\text{L}$ read
% \begin{equation}
% \Delta\vec{v}_1^{\,\prime}=
% -\frac{m_2}{m_1+m_2}
% \left(
% \dot{r}^{\prime}\vec{e}_r^{\,\prime}+
% r^{\prime}\vec{e}_{\dot{r}}^{\,\prime}
% \right),
% \end{equation}
%. The corresponding vector for particle 2 reads
% \begin{equation}
% \Delta\vec{v}_2^{\prime}=
% \frac{m_1}{m_1+m_2}
% \left(
% \dot{r}^{\prime}\vec{e}_r^{\prime}+
% r^{\prime}\vec{e}_{\dot{r}}^{\prime}
% \right),
% \end{equation}
\begin{equation}
  \label{eq:2}
\left(
  \begin{array}{c}
\Delta\vec{v}_1^{\,\prime} \\    
\Delta\vec{v}_2^{\,\prime}
  \end{array}
\right)=
\left(
\begin{array}{r}
-m_2\\
m_1  
\end{array}
\right)
\frac{1}{m_1+m_2}
\left(
\dot{r}^{\prime}\vec{e}_r^{\,\prime}+
r^{\prime}\vec{e}_{\dot{r}}^{\,\prime}
\right)\:.
\end{equation}
With this, the post-collisional velocities expressed in the laboratory frame read
\begin{equation}
\label{eq:post-velocities}
\vec{v}_i^{\,\prime}=
\dot{\vec{R}}^{\prime}+
\Delta\vec{v}_i^{\,\prime},
\end{equation}
%which completes the collision mapping. Where, i
In absence of external fields we have $\dot{\vec{R}}^{\prime}=\dot{\vec{R}}^0$. 

Together with the matrix of restitution, Eq.~\eqref{eq:colMat}, Equations \eqref{eq:post-positions} and \eqref{eq:post-velocities} establish a complete set of equations for the computation of the post-collisional positions and velocities from the pre-collisional values.
 
\section{Collision of Granular Particles}
\label{sec:lookup}

The previous section provides a general way to perform event-driven simulations of soft particles, that is, the hard-sphere approximation, Eq. \eqref{eq:fixedPos} is not exploited. So far, however, we did not specify the particle interaction force which determines the properties of the matrix of restitution, Eq. \eqref{eq:colMat}.

In this section we consider two widely used models for the interaction force $F_n$, the linear dashpot model and the model of viscoelastic spheres to obtain the matrix of restitution, \eqRef{eq:colMat}. Both models are characterized by many material and system parameters, thus, the components of the matrix of restitution are functions of these parameters. Since, for both force models, an analytical evaluation is not possible, by appropriate scaling we reduce the problem to three independent parameters, leading to a convenient way for computing efficient lookup tables for the matrix of restitution. 

Together with the collision rule, Eqs. \eqref{eq:post-positions} and \eqref{eq:post-velocities}, the results of this section allow for highly efficient event-driven simulation of granular gases of \emph{soft} spheres.
% and might direct a way to a kinetic theory of \emph{soft} sphere granular gases. 

\subsection{Linear-Dashpot Model}
\label{sec:colMapLinDash}
The linear-dashpot model is widely used in the literature for the simulation of granular systems. Its physical relevance may be questioned since neither the elastic \cite{hertz1881} nor the dissipative part of the force \cite{brilliantov1996} agree with physical reality. It even violates a dimension analysis \cite{ramirez1999}. Its main characteristics is that in the hard-sphere limit it leads to a coefficient of restitution which is independent of the impact velocity (which disagrees with experiments as well, e.g. \cite{Bridges}). Although physically questionable, the linear-dashpot model is widely used since its consequence, the constant coefficient of restitution, simplifies the analytical analysis largely. Therefore, except for very few examples, e.g. \cite{Brilliantov:2005,Brilliantov:2004,Brilliantov:2003,Brilliantov:2002}, virtually the entire Kinetic Theory of granular gases relies on this assumption. 

The linear-dashpot model defines the normal force between colliding spheres by
\begin{equation}
\label{eq:fNLinDash}
 F_n=k(l-r)-\gamma\dot{r}\,,
\end{equation}
with $l\equiv R_1+R_2$, and $k$ and $\gamma$ being the spring constant and the dissipative parameter.  With this force and the scaling (see \eqRef{eq:scaleDef})
\begin{equation}
\label{eq:scalingLinDash}
 \varPhi\equiv1,\quad T\equiv\frac{1}{\omega},\quad X\equiv\frac{\dot{r}(0)}{\omega},\quad\omega\equiv\sqrt{\frac{k}{m_\text{eff}}}\:,
\end{equation}
from \eqRef{eq:newtonPolarScaled} we obtain the equations of motion
\begin{equation}
\label{eq:newtonLinDashScaled}
\begin{split}
\frac{\mathrm d \tilde{\varphi}}{\mathrm d
\tilde{t}}&=\frac{c_\varphi}{\tilde{r}^2}\\
\frac{\mathrm d^2\tilde{r}}{\mathrm d\tilde{t}^2}&=
\frac{c_\varphi^2}{\tilde{r}^3}+\left(\tilde{l}-\tilde{r}\right)-c_\text{dis}\frac{\mathrm d\tilde{r}}{\mathrm d\tilde{t}}\:,
\end{split}
\end{equation}
where 
\begin{equation}
  \label{eq:ltilde-cdis}
  \tilde{l}\equiv \frac{l}{X}\,,~~~~~c_\text{dis}\equiv\frac{\gamma\,T}{m_\text{eff}}\,.   
\end{equation}

We solve \eqRef{eq:newtonLinDashScaled} with the initial conditions (see \eqRef{eq:initCondScaled})
\begin{equation}
\label{eq:initCondLinDashScaled}
 \tilde{\varphi}(0)=0,\quad\tilde{r}(0)=\tilde{l}\quad\text{and}\quad\frac{\mathrm d \tilde{r}}{\mathrm d \tilde{t}}(0)=-1
\end{equation}
for a given set of $\{\tilde{l},c_\varphi,c_\text{dis}\}$ in the interval $0\le\tilde{t}\le \tilde{\tau}$, where $\tilde{\tau}$ is the time where the collision ceases given by the condition \eqRef{eq:tauDef}. The matrix of restitution, \eqRef{eq:colMat}, is then obtained by using the definitions of its components, Eqs. \eqref{eq:epsCompDefA}-%\eqRef{eq:epsCompDefB}, 
\eqref{eq:epsCompDefC}, \eqref{eq:epsCompDefD} and \eqref{eq:epsCompDefE}.

\begin{figure}[h!]
 \includegraphics[width=0.8\columnwidth,clip]{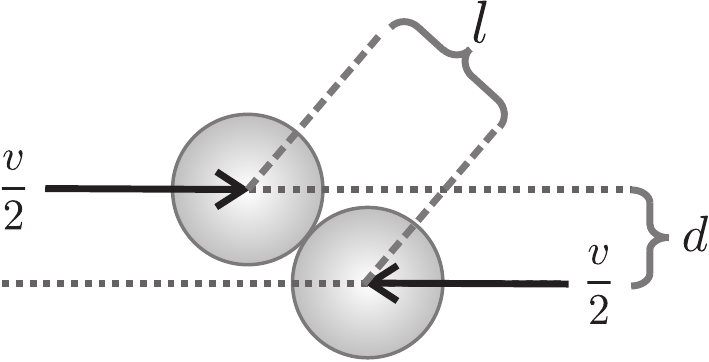}
 \caption{Eccentric collision of spheres.}
 \label{fig:excCollSetup}
\end{figure}

The reduced set of parameters, $\{\tilde{l},c_\varphi,c_\text{dis}\}$, follows from both, material parameters ($k$, $\gamma$, mass density $\rho$), particle sizes ($R_1$, $R_2$) and impact parameters (impact velocity $v$ and eccentricity $e\equiv d/l$, see \figRef{fig:excCollSetup}). 

\begin{figure}[h!]
\includegraphics[width=\figWidth,viewport=0 0 1190 800,clip]{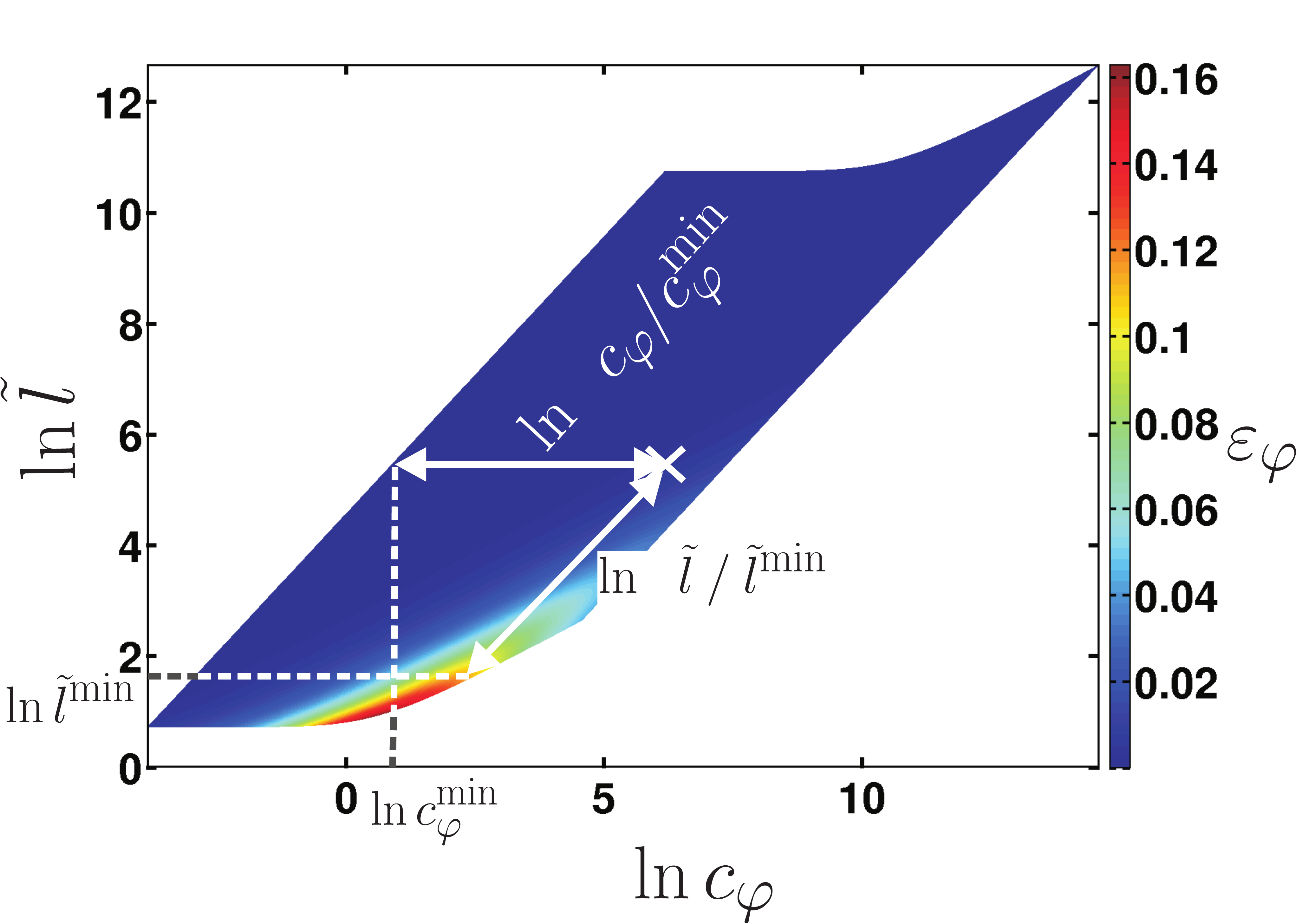}
\caption{The space of physical parameters given in Tab. \ref{tab:physParamSpaceLinDash} translates into a space of the scaled parameters $\left(c_\varphi/c_\varphi^\text{min}, \tilde{l}/\tilde{l}^\text{min}\right)$ of complex shape. The figure shows one of the elements of the matrix of restitution, $\varepsilon_\varphi$, as a function of the scaled variables, for the special case $\gamma=0$ (elastic collisions). Each point of the colored region corresponds to a point in the physical space given in Tab. \ref{tab:physParamSpaceLinDash}. The white regions are inaccessible within the chosen set of physical parameters. Expressions for $c_\varphi^\text{min}$ and $\tilde{l}^\text{min}$ are given in Eqs. \eqref{eq:cPhiDef} and \eqref{eq:ltilde-cdis}.}
\label{fig:illustRectLinDash}
\end{figure}

For practical application, we need the matrix of restitution, \eqRef{eq:colMat}, for a wide range of the (physical) system parameters, corresponding to a certain area in the $\{\tilde{l},c_\varphi,c_\text{dis}\}$-space of complicated shape. For elastic spheres ($\gamma=0$), the intervals for the physical parameters given in \tabRef{tab:physParamSpaceLinDash} lead to the area in the  $\{\tilde{l},c_\varphi,c_\text{dis}\}$-space shown in \figRef{fig:illustRectLinDash} showing $\varepsilon_\varphi$ as a function of $\tilde{l}$ and $c_\varphi$.
\begin{table}[h!]
\centering
\begin{tabular}{l@{~~~~}l@{~~~~}l@{~~~~}l@{~~~~}l}
\hline\hline
&unit&min.&max.&\\%\hline
$k$&[$10^3$ N$/$m]&$1$&$1000$&spring constant\\%\hline
$R$&[m]&$0.001$&$0.1$&particle radius\\%\hline
$\rho_m$&[kg$/$m$^3$]&$250$&$3250$&material density\\%\hline
$\gamma$&[kg$/$s]&$0.01$&$1.25$&dissipative parameter\\%\hline
$v$&[m$/$s]&$0.001$&$25$&impact velocity\\%\hline
$d/l$&-&$0.01$&$0.99$&eccentricity\\\hline
\end{tabular}

\caption{Space of physical parameters used to obtain the matrix of restitution for the linear-dashpot model.
%\figRef{fig:colMapLinDash}. 
For the definition of impact velocity and eccentricity see \figRef{fig:excCollSetup}.}
\label{tab:physParamSpaceLinDash}
\end{table}

We switch now from $(\tilde{l}, c_\varphi, c_\text{dis})$ to a new set of independent parameters, such that the parameter space is bound by perpendicular straight axis. This is necessary for the numerically efficient access to the elements of the matrix of restitution (represented as a lookup table) needed for efficient eMD simulations. 

From the definitions \eqRef{eq:cPhiDef}, \eqRef{eq:angMomDef}, $X\equiv l/\tilde{l}$ and the geometry of the collision, \figRef{fig:excCollSetup}, we find
\begin{equation}
\label{eq:cPhilTildeRelLinDash}
 \ln c_\varphi=\ln\tilde{l} -\frac{1}{2}\ln  \left(\frac{1}{e^2}-1\right)\:,
\end{equation}
which indicates that for a given impact eccentricity, $e$, all possible $\{\ln \tilde{l}, \ln c_\phi\}$-pairs are located on a straight line of slope $1$ \cite{mueller2011} with $-4.6\lesssim -\frac{1}{2}\ln  \left(\frac{1}{e^2}-1\right) \lesssim 1.95$ for the parameters given in \tabRef{tab:physParamSpaceLinDash}. That is, for a given $\tilde{l}$ the smallest accessible $c_\varphi$ is given by
\begin{equation}
\ln c_\varphi^\text{min}\equiv\ln\tilde{l} -\frac{1}{2}\ln  \left(\frac{1}{\left(e^\text{min}\right)^2}-1\right)=\ln\tilde{l}+g^\text{min}
\end{equation}
with $g^\text{min} \approx -4.6$.
By switching to $\ln \frac{c_\varphi}{c_\varphi^\text{min}}$, the lines of constant eccentricity $e$ in the $\{\ln \tilde{l}, \ln c_\phi\}$-space are hence raised to straight vertical lines (see \figRef{fig:illustRectLinDash}).

Further, from the definition of $\tilde{l}$, Eq. \eqref{eq:ltilde-cdis}, the scaling \eqRef{eq:scalingLinDash} and geometry, we obtain
\begin{equation}
 \tilde{l}=\frac{l\omega}{v\sqrt{1-e^2}}\:.
\end{equation}
Using \eqRef{eq:cPhilTildeRelLinDash},  we express $e$ in terms of $\tilde{l}\text{ and }c_\phi$ and end up with
\begin{equation}
\label{eq:lTildeRelatLinDash}
 \ln \tilde{l}=\ln \left(\frac{l\omega}{v}\right)+\frac{1}{2}\ln\left[1+\left(\frac{c_\varphi}{\tilde{l}}\right)^2\right]\:.
\end{equation}

For the physical parameters in \tabRef{tab:physParamSpaceLinDash}  we obtain $-2.3 \lesssim \ln \left(\frac{l\omega}{v}\right) \lesssim 14.83$, thus, for a given impact eccentricity, the smallest attainable $\tilde{l}$ is hence given by
\begin{equation}
  \ln \tilde{l}^\text{min}=m^\text{min}+\frac{1}{2}\ln\left[1+\left(\frac{c_\varphi}{\tilde{l}}\right)^2\right]\:,
\end{equation}
with $m^\text{min}\approx -2.3$.

Consequently, if we would plot \figRef{fig:illustRectLinDash} with axis $\ln \nicefrac{c_\varphi}{c_\varphi^\text{min}}$ (instead of  $\ln c_\varphi$) and $\ln \nicefrac{\tilde{l}}{\tilde{l}^\text{min}}$ (instead of $\ln \tilde{l}$), the accessible data points would form a rectangular area. Thus, the complicated shaped colored region in \figRef{fig:illustRectLinDash} is transformed into a rectangle which allows for an efficient use of a corresponding lookup table in eMD simulations. The generalization to inelastic particles is straightforward.

The inverse transformation from $\{\ln \nicefrac{c_\varphi}{c_\varphi^\text{min}},\ln \nicefrac{\tilde{l}}{\tilde{l}^\text{min}}\}$ to $\{\ln c_\varphi,\ln \tilde{l}\}$ is obtained directly from the definitions of $\ln \nicefrac{c_\varphi}{c_\varphi^\text{min}}$ and $\ln \nicefrac{\tilde{l}}{\tilde{l}^\text{min}}$:
\begin{equation}
  \label{eq:3}
  \begin{split}
	\ln\tilde{l}&=\ln \frac{\tilde{l}}{\tilde{l}^\text{min}}
					+m^\text{min}
					+\frac{1}{2}\ln\left[
						1+e^{2\left(\ln \frac{c_\varphi}{c_\varphi^\text{min}}+g^\text{min}\right)}
					\right]
	\\
	\ln c_\varphi&=\frac{c_\varphi}{c_\varphi^\text{min}}
						+\ln\tilde{l}+g^\text{min}\:.    
  \end{split}
\end{equation}

% \begin{align}
% 	\ln\tilde{l}&=
% 					\ln \nicefrac{\tilde{l}}{\tilde{l}^\text{min}}
% 					+m^\text{min}
% 					+\frac{1}{2}\ln\left[
% 						1+e^{2(\ln \nicefrac{c_\varphi}{c_\varphi^\text{min}}+g^\text{min})}
% 					\right]
% 	\\\nonumber
% 	\ln c_\varphi&=\nicefrac{c_\varphi}{c_\varphi^\text{min}}
% 						+\ln\tilde{l}+g^\text{min}\:.
% \end{align}

\begin{figure*}[ht]
\centering
 \includegraphics[width=0.99\textwidth]{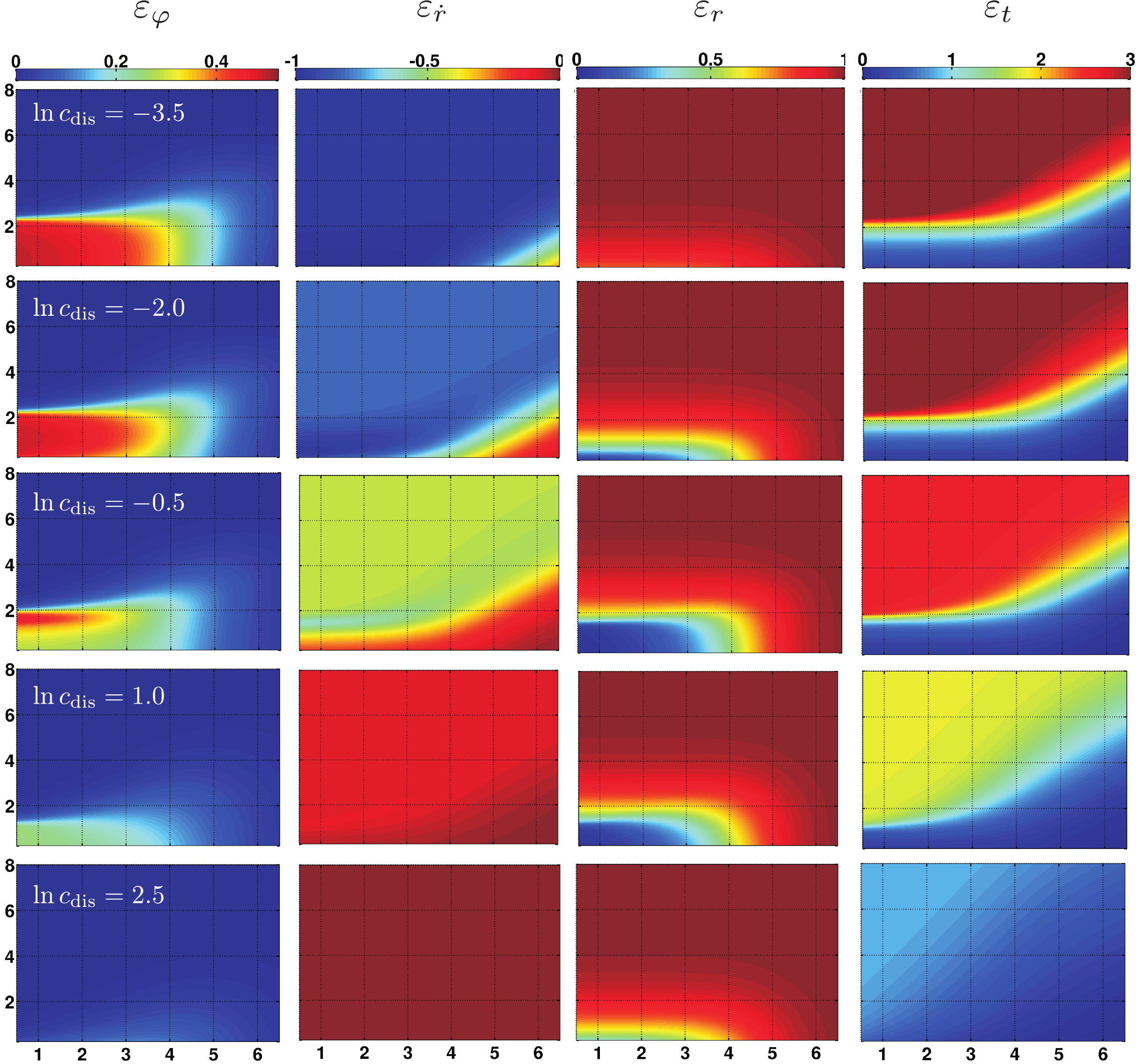}
\caption{(color online) Components of the matrix of restitution, \eqRef{eq:colMat}, for the linear-dashpot interaction force, \eqRef{eq:fNLinDash} and some values of $\ln c_\text{dis}$. Abscissa of all panels: $\ln\nicefrac{c_\varphi}{c_\varphi^\text{min}}$. Ordinate of all panels: $\ln\nicefrac{\tilde{l}}{\tilde{l}^\text{min}}$. 
%Each row displays $\varepsilon_\varphi$, $\varepsilon_{\dot{r}}$, $\varepsilon_r$ and $\varepsilon_t$ for the dissipative parameter $\ln c_\text{dis}$ indicated by the white label in the corresponding image for $\varepsilon_\varphi$. 
The range of scaled parameters $(c_\varphi, \tilde{l})$ corresponds to the physical parameters space defined in \tabRef{tab:physParamSpaceLinDash}.}
 \label{fig:colMapLinDash}
\end{figure*}

Figure \ref{fig:colMapLinDash} shows the components of the matrix of restitution, \eqRef{eq:colMat}, for the linear dashpot interaction force, \eqRef{eq:fNLinDash} and the range of physical parameters specified in \tabRef{tab:physParamSpaceLinDash}. Each row corresponds to a certain (scaled) dissipative constant $\ln c_\text{dis}$, see Eq. \ref{eq:ltilde-cdis}. 
%\begin{widetext}
%\end{widetext}
%\clearpage

The first column of \figRef{fig:colMapLinDash} displays $\varepsilon_\varphi$ (see \eqRef{eq:epsCompDefB}) which describes the rotation of the inter-particle unit vector $\vec{e}_r$ during the contact. This rotation angel is determined by the contact duration, $\tau$, and the rotation velocity, $\dot{\varphi}$. If we would disregard centrifugal forces, the rotation velocity would be constant. As the contact duration decreases with inelasticity the rotation angle and, thus, $\varepsilon_\varphi$ also decrease with inelasticity. Regarding the component $\varepsilon_\varphi$, elastic collisions hence represent the marginal case \cite{mueller2011}.

The second column in \figRef{fig:colMapLinDash} (see Eq. \eqref{eq:epsCompDefE}) shows $\varepsilon_{\dot{r}}$. The coefficient $\varepsilon_n=-\varepsilon_{\dot{r}}$ is the well known coefficient of normal restitution including effects due to centrifugal forces. It describes the loss of energy of the particles' relative velocity in normal direction, due to the collision. From this interpretation follows that $-\varepsilon_{\dot{r}}$ decreases with increasing dissipation for all combinations of $\{\ln c_\varphi/c_\varphi^\text{min},\ln \tilde{l}/\tilde{l}^\text{min}\}$. 

The coefficient $\varepsilon_r$ (see Eq. \eqref{eq:epsCompDefA}), shown in the third column of \figRef{fig:colMapLinDash} stands for the ratio of the post- and precollisional distance of the particles. Due to the premature end of collision (\eqRef{eq:tauDef}, see \cite{schwager2007,schwager2008} for an in-depth discussion) the value of $\varepsilon_r$ may differ from 1 for inelastic collisions ($\gamma>0$). For impacts leading to large rotation velocity, $\dot{\varphi}$, the coefficient $\varepsilon_r$ may significantly deviate from unity because of centrifugal forces. While dissipative forces cause a premature and of the collision, they also reduce the rotation velocity. Consequently there is an optimal value for the damping coefficient, $\gamma$ (or its scaled value $c_\text{dis}$), which minimizes $\varepsilon_r$.
% (see third column of \figRef{fig:colMapLinDash}).

The last column of \figRef{fig:colMapLinDash} shows the component $\varepsilon_t$ (see Eq. \eqref{eq:epsCompDefD}) which stands for the collision duration measured in units of the characteristic time $T$ (see \eqRef{eq:scaleDef}).
% Independent of damping, there is a characteristic distribution of $\varepsilon_t$ within the $\{\ln c_\varphi/c_\varphi^\text{min},\ln \tilde{l}/\tilde{l}^\text{min}\}$-plain. 
The absolute value of $\varepsilon_t$ decreases with damping since due to the premature end of collision, the contact duration, $\tau$, decreases with increasing dissipation.

\subsection{Viscoelastic Spheres}
The normal component of the interaction force between two colliding viscoelastic spheres reads
\begin{equation}
\label{eq:fNViscel}
F_n=F_n^{\text{el}}+ F_n^{\text{dis}} =\rho_\text{el}(l-r)^{\nicefrac{3}{2}} -\frac{3}{2}A\rho_\text{el}\dot{r}\sqrt{l-r}\:,
\end{equation}
where
\begin{equation}
  \label{eq:rhodef}
 \rho_\text{el}\equiv\frac{2Y\sqrt{R_{\text{eff}}}}{3(1-\nu^2)}
\end{equation}
and $Y$, $\nu$ and $R_{\text{eff}}$ denote the Young modulus, the Poisson ratio and the effective radius $R_{\text{eff}}=R_1R_2/(R_1+R_2)$, respectively. The elastic part $F_n^{\text{el}}$ of this widely used collision model \cite{kruggelEmden2007,stevens2005,schaefer1996} is given by the Hertz contact force \cite{hertz1881}. The dissipative part, $F_n^{\text{dis}}$, was first motivated in \cite{kuwabara1987} and then rigorously derived in \cite{brilliantov1996} and \cite{morgado1997}, where only the approach in \cite{brilliantov1996} leads to an analytic expression for the parameter $A$, being a function of the elastic and viscous material parameters, see \cite{brilliantov1996} for details.

Using the normal force \eqRef{eq:fNViscel} and the scaling relation Eq. \eqref{eq:scaleDef} with
\begin{equation}
 \varPhi\equiv1,\quad T\equiv\frac{1}{k^{\nicefrac{2}{5}}\left(-\dot{r}^0\right)^{\nicefrac{1}{5}}},\quad X\equiv\frac{\left(-\dot{r}^0\right)^{\nicefrac{4}{5}}}{k^{\nicefrac{2}{5}}}\:,
\end{equation}
where $k\equiv \rho / m_\text{eff}$, the general equation of motion \eqRef{eq:newtonPolarScaled} reads
\begin{equation}
\label{eq:newtonViscelScaled}
\begin{split}
\frac{\mathrm d \tilde{\varphi}}{\mathrm d
\tilde{t}}&=\frac{c_\varphi}{\tilde{r}^2}\\
 \frac{\mathrm d^2\tilde{r}}{\mathrm d\tilde{t}^2}&=
 \frac{c_\varphi^2}{\tilde{r}^3}+\left(\tilde{l}-\tilde{r}\right)^{\nicefrac{3}{2}}-c_\text{dis}\frac{\mathrm d\tilde{r}}{\mathrm d\tilde{t}}\sqrt{\tilde{l}-\tilde{r}}\:,
\end{split}
\end{equation}
where $c_\text{dis}\equiv\frac{3A}{2T}$.

Proceeding along the lines of \secRef{sec:colMapLinDash}, we solve \eqRef{eq:newtonViscelScaled} with the initial conditions \eqRef{eq:initCondLinDashScaled} for a given range of physical parameters to obtain the matrix of restitution, \eqRef{eq:colMat}. The intervals of parameters specified in Table \ref{tab:physParamSpaceViscel} cover a wide range of applications.
\begin{table}[h!]
\centering
\begin{tabular}{l@{~~~}l@{~~~}l@{~~~}l@{~~~}l}
\hline\hline
&unit&min.&max.&\\%\hline
$Y$&[$10^9$ N$/$m$^2$]&$0.01$&$100$&Young's Modulus\\%\hline
$\nu$&-&$0.2$&$0.5$&Poisson's ratio\\%\hline
$R$&[m]&$0.001$&$0.1$&particle radius\\%\hline
$\rho_m$&[kg$/$m$^3$]&$250$&$3250$&material density\\%\hline
$A$&[s]&$10^{-6}$&$1$&dissipative parameter\\%\hline
$v$&[m$/$s]&$0.001$&$25$&impact velocity\\%\hline
$d/l$&-&$0.01$&$0.99$&eccentricity\\\hline
\end{tabular}
\caption{Parameter space scanned to obtain the matrix of restitution for viscoelastic spheres.
%\figRef{fig:colMapViscel}. 
For the definition of impact velocity and eccentricity see \figRef{fig:excCollSetup}.}
\label{tab:physParamSpaceViscel}
\end{table}

The set of physical parameters can be transformed in a set of scaled variables, $\{\tilde{l},c_\varphi,c_\text{dis}\}$. Again, the specified ranges of physical parameters correspond to a region in the $\{\tilde{l},c_\varphi,c_\text{dis}\}$-space of complicated shape. As in the case of the linear-dashpot model we look for a transformation such that the admitted sets of parameters establish a rectangular system.

The parameters of the force do not enter \eqRef{eq:cPhilTildeRelLinDash}, therefore, it holds true for viscoelastic spheres too. Since the marginal values for the impact eccentricity, $e$, remain (same ranges in Tabs. \ref{tab:physParamSpaceLinDash} and \ref{tab:physParamSpaceViscel}), again $g^\text{min}=-4.6$.

The corresponding equation to  Eq. \eqref{eq:lTildeRelatLinDash} valid for the linear-dashpot model, reads

%Equation \eqref{eq:lTildeRelatLinDash} for the linear-dashpot model reads \todo{for the case of viscoelastic spheres:}
\begin{equation}
 \ln\tilde{l}=
					\ln\left[l\left(\frac{k}{v^2}\right)^{\nicefrac{2}{5}}\right]
					+\frac{2}{5}\ln\left[1+\left(\frac{c_\varphi}{\tilde{l}}\right)^2\right]
\end{equation}
for the case of viscoelastic spheres.
%\todo{for the case of viscoelastic spheres.}
%\patric{Nach meinem Empfinden sollten wir das erste todo streichen und das zweite uebernehmen.}
Using the parameters from Tab. \ref{tab:physParamSpaceViscel} we obtain for the first term
\begin{equation}
  \label{eq:4}
0.75 \approx m^\text{min} < \ln\left[l\left(\frac{k}{v^2}\right)^{\nicefrac{2}{5}}\right] <  m^\text{max}\approx13.66  
\end{equation}
With this we define
\begin{equation}
  \ln \tilde{l}^\text{min}=m^\text{min}+\frac{2}{5}\ln\left[1+\left(\frac{c_\varphi}{\tilde{l}}\right)^2\right]\:.
\end{equation}

In the same way as for the linear-dashpot force, we use $\left[\ln \left(c_\varphi/c_\varphi^\text{min}\right),~\ln \left(\tilde{l}/\tilde{l}^\text{min}\right)\right]$ 
instead of  $\left[\ln c_\varphi\,,~\ln \tilde{l}\right]$ as independent variables. While the domain of physical parameters (Tab. \ref{tab:physParamSpaceViscel}) is represented by an area of complex shape in the coordinates $\left[\ln c_\varphi\,,~\ln \tilde{l}\right]$ (similar to \figRef{fig:illustRectLinDash}), in the new variables, the domain is bound by a rectangle which is much better suited for the construction of a lookup table for the matrix of restitution. 

In contrast to the linear dashpot model discussed in \secRef{sec:colMapLinDash}, for viscoelastic spheres the dissipative parameter $c_\text{dis}=\frac{3A}{2T}$ depends on $\tilde{l}$ and $c_\varphi$ via $T$. From the definitions of $c_\varphi$, $\tilde{l}$, $X$, $T$, $\vec{L}$ and geometry, we obtain
\begin{equation}
T=\frac{lX}{dv}\frac{c_\varphi}{\tilde{l}}=\frac{l}{v}\frac{l}{d}\frac{c_\varphi}{\tilde{l}^2}\:.
\end{equation}
Using \eqRef{eq:cPhilTildeRelLinDash} to replace $\nicefrac{l}{d}=\nicefrac{1}{e}$ and the definition of $c_\text{dis}$ (below \eqRef{eq:newtonViscelScaled}) this yields
\begin{equation}
 \ln c_\text{dis}=\ln\left(\frac{3}{2}A\frac{v}{l}\right)
						+\ln\tilde{l}
						-\frac{1}{2}\ln\left[1+\left(\frac{c_\varphi}{\tilde{l}}\right)^2\right]\:.
\end{equation}
That is, for a given $c\equiv\ln\left(\frac{3}{2}A\frac{v}{l}\right)$, $\ln c_\text{dis}(\tilde{l},c_\varphi)$ forms a curved surface in the $\{\tilde{l},c_\varphi,c_\text{dis}\}$-space,
where $c$ ranges from $c^\text{min}\approx-18.71$ to $c^\text{max}\approx9.84$ for the physical parameters given in \tabRef{tab:physParamSpaceViscel}. With this, we define 
\begin{equation}
 \ln c_\text{dis}^\text{min}=c^\text{min}
						+\ln\tilde{l}
						-\frac{1}{2}\ln\left[1+\left(\frac{c_\varphi}{\tilde{l}}\right)^2\right]\:.
\end{equation}
Using $\ln \nicefrac{c_\text{dis}}{c_\text{dis}^\text{min}}$ instead of $\ln c_\text{dis}$, the physical parameters given in \tabRef{tab:physParamSpaceViscel} are mapped to a cube-shaped domain in the $\ln \nicefrac{c_\varphi}{c_\varphi^\text{min}}$-$\ln \nicefrac{\tilde{l}}{\tilde{l}^\text{min}}$-$\ln \nicefrac{c_\text{dis}}{c_\text{dis}^\text{min}}$-space, allowing for efficient lookup tables.

\begin{figure*}[h!]
 \centering
  \includegraphics[width=0.99\textwidth]{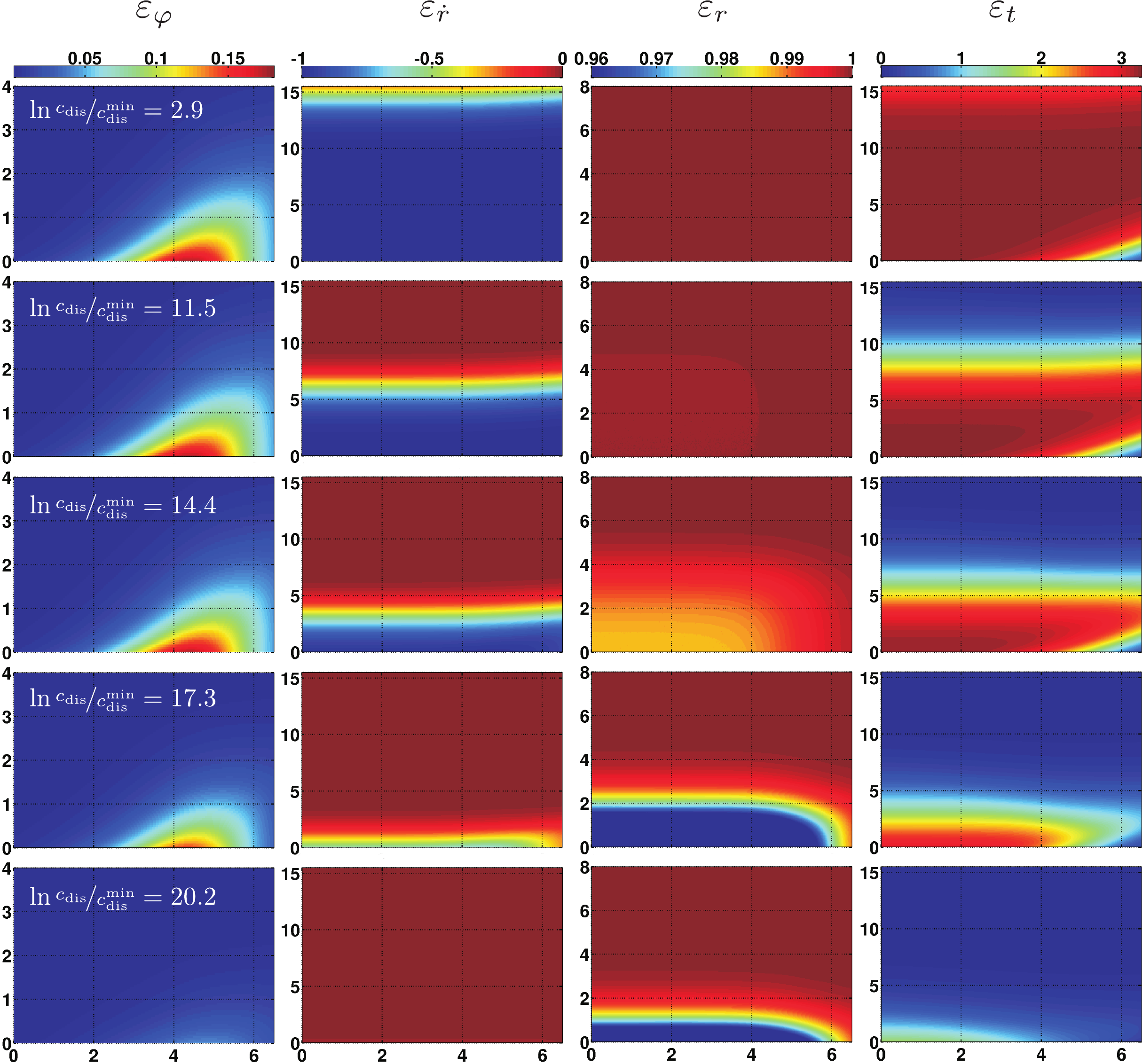}
  \caption{Components of the collision mapping \eqRef{eq:colMat} for the viscoelastic interaction \eqRef{eq:fNViscel}. Abscissa of all panels: $\ln\nicefrac{c_\varphi}{c_\varphi^\text{min}}$. Ordinate of all panels: $\ln\nicefrac{\tilde{l}}{\tilde{l}^\text{min}}$. Each row displays $\varepsilon_\varphi$, $\varepsilon_{\dot{r}}$, $\varepsilon_r$ and $\varepsilon_t$ for the dissipative parameter $\ln\nicefrac{c_\text{dis}}{c_\text{dis}^\text{min}}$ indicated by the white label in the corresponding image for $\varepsilon_\varphi$. Parameters as indicated in \tabRef{tab:physParamSpaceViscel} (color online).}
  \label{fig:colMapViscel}
 \end{figure*}
 %\clearpage

\figRef{fig:colMapViscel} displays the result for a selection of dissipative parameters $\ln\nicefrac{c_\text{dis}}{c_\text{dis}^\text{min}}$. Similar to \figRef{fig:colMapLinDash} each row of \figRef{fig:colMapViscel} shows the four components of the collision mapping \eqRef{eq:colMat} for a fixed dissipative parameter $\ln\nicefrac{c_\text{dis}}{c_\text{dis}^\text{min}}$. Again, dissipation increases from the top to the bottom row.  The discussion of \figRef{fig:colMapViscel} is absolutely equivalent to the linear dashpot case, \figRef{fig:colMapLinDash}. 

The corresponding transformation back to $\tilde{l}$, $c_\varphi$ and $c_\text{dis}$ may be obtained directly from the definitions and reads
\begin{equation}
  \label{eq:5}
  \begin{split}
    \ln\tilde{l}&=\ln \frac{\tilde{l}}{\tilde{l}^\text{min}}+m^\text{min} +\frac{2}{5}\ln\left[1+e^{2\left(\ln \frac{c_\varphi}{c_\varphi^\text{min}}+g^\text{min}\right)}\right]\\
    \ln c_\varphi&=\frac{c_\varphi}{c_\varphi^\text{min}}+\ln\tilde{l}+g^\text{min}\\
    \ln c_\text{dis}&=\ln \frac{c_\text{dis}}{c_\text{dis}^\text{min}}+c^\text{min}+\ln\tilde{l}-\frac{1}{2}\ln\left[1+e^{2\left(\ln \frac{c_\varphi}{c_\varphi^\text{min}}+g^\text{min}\right)}\right]
  \end{split}
\end{equation}
% \begin{align}
% 	\ln\tilde{l}&=
% 					\ln \nicefrac{\tilde{l}}{\tilde{l}^\text{min}}
% 					+m^\text{min}\\\nonumber
% 					&+\frac{2}{5}\ln\left[
% 						1+e^{2(\ln \nicefrac{c_\varphi}{c_\varphi^\text{min}}+g^\text{min})}
% 					\right]\\\nonumber
% 	\ln c_\varphi&=\nicefrac{c_\varphi}{c_\varphi^\text{min}}
% 						+\ln\tilde{l}+g^\text{min}\\\nonumber
% 	\ln c_\text{dis}&=
% 					\ln \nicefrac{c_\text{dis}}{c_\text{dis}^\text{min}}
% 					+c^\text{min}
% 					+\ln\tilde{l}\\\nonumber
% 					&-\frac{1}{2}\ln\left[
% 						1+e^{2(\ln \nicefrac{c_\varphi}{c_\varphi^\text{min}}+g^\text{min})}
% 					\right]\:.
% \end{align}

\section{Event-driven Molecular Dynamics Algorithm}

\subsection{Traditional event-driven Molecular Dynamics}
\label{sec:trad-event-driv}

The traditional eMD scheme of {\em hard} particles is rather simple although an efficient implementation allowing for the simulation of many millions of particles may be technically rather complex, see e.g. \cite{Bannerman}. Its basic concept is to
\begin{enumerate}
\item find the next colliding pair $(i,j)$ of particles in the system and their collision time $t^*$
\item propagate all particles $k$ to this time,
  \begin{equation}
    \vec{r}_k := \vec{r}_k+\vec{v}_k(t^*-t)
 \end{equation}
where $t$ is the present time and $\vec{v}_k$ is the present velocity of particle $k$
\item compute the post-collisional velocities of particles $i$ and $j$ due to the collision rule
\begin{equation}
\label{eq:cRuleHS}
\begin{split}
  \vec{v}_i &:= \vec{v}_i+\frac{1+\varepsilon_n}{2}\:\left[\left(\vec{v}_i-\vec{v}_j\right)\cdot\vec{e}_r\right] \vec{e}_r\\
  \vec{v}_j &:= \vec{v}_j-\frac{1+\varepsilon_n}{2}\:\left[\left(\vec{v}_i-\vec{v}_j\right)\cdot\vec{e}_r\right] \vec{e}_r
% \dot{\vec{r}}_2^{\,\prime} &=\dot{\vec{r}}_2^0-\frac{1+\varepsilon_n^\text{HS}}{2}\:\dot{\vec{r}}\,\vec{e}_r
\end{split}
\end{equation}
where $\varepsilon_n$ is the coefficient of normal restitution. For simplicity of the notation we consider here particles of identical mass, the generalization is straightforward.
\item continue with step 1
\end{enumerate}

While all eMD schemes work in principle as described, there are many ways to increase the efficiency, e.g. \cite{lubachevsky1991,marin1993,poeschel2005,Bannerman}, which shall not be discussed here. Moreover, the scheme described above does not take into account external fields like gravity, interaction with (moving) boundaries etc.

\begin{figure}[ht]
 \includegraphics[width=0.9\columnwidth,clip]{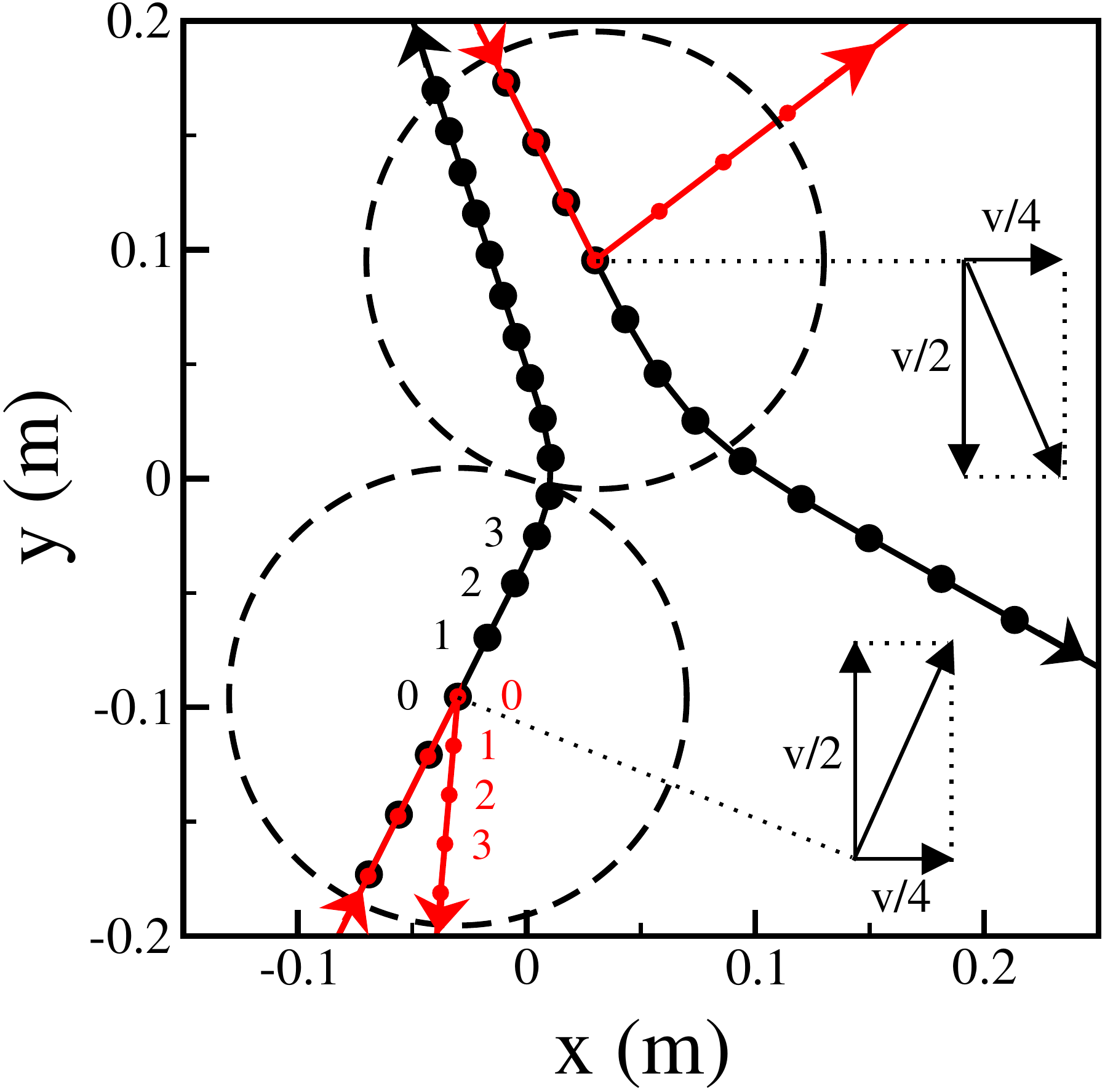}\\
\includegraphics[width=0.9\columnwidth,clip]{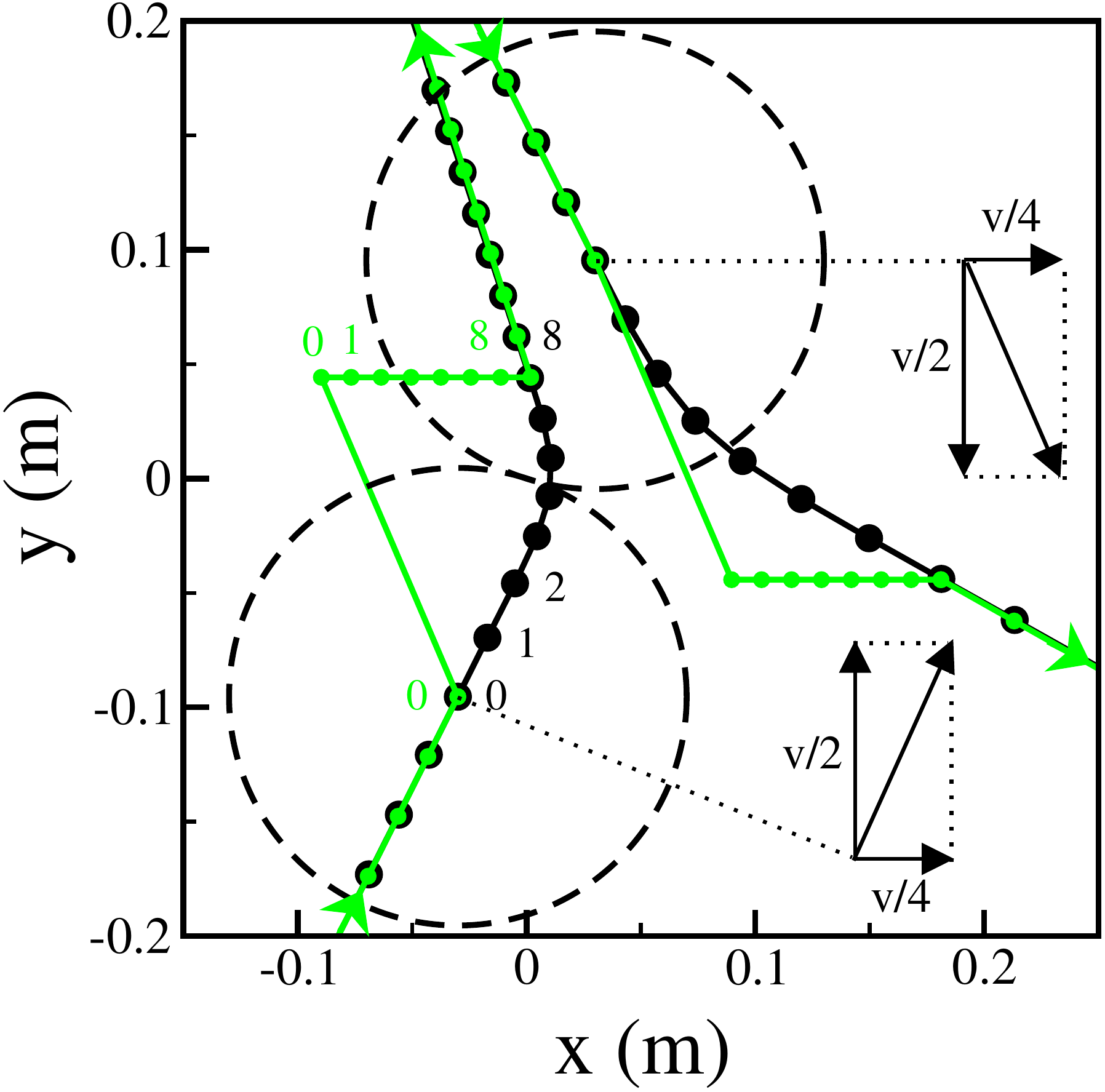}
\caption{(color online) Traces of two colliding spheres (parameters are given in the text). Black lines show the numerical integration of Newton's equation (MD), red lines (top) show the trajectories as obtained from eMD with the assumption of instantaneous collisions. The green lines (bottom) show the trajectories as obtained by the new eMD algorithm (see \secRef{sec:impr-event-driv}). Symbols and numbers (of the respective color) indicate the particle positions at equidistant points in time. The number 0 stands for the moment when the particles touch and 7 corresponds to the end of the collision (stepsize $dt=\tau/7$). The dashed circles show the spheres at the moment of impact.}
 \label{fig:illustrateCRule}
\end{figure}

Figure \ref{fig:illustrateCRule} (top) shows the trajectories of two colliding particles as obtained by eMD in comparison to force-based MD, that is, the numerical solution of Newton's equation of motion. For the interaction force we assume a linear-dashpot model, Eq.~\eqref{eq:fNLinDash}, with the parameters $k=2~\text{kN}/\text{m}$, $R=0.1~\text{m}$, $\rho=1140~\text{kg}/\text{m}^3$, $v=5~\text{m}/{s}$, $e=0.3$ (see \figRef{fig:excCollSetup}). Initial velocities are $\vec{v}_1=(v/4,\,v/2,0)$ and   $\vec{v}_2=(v/4,\, -v/2,0)$. Since here we assume elastic interaction ($\gamma=0$), the corresponding coefficient of normal restitution is $\varepsilon_n=1$.

%\clearpage
 
The figure reveals two fundamental problems which are both attributed to the assumption of instantaneous collisions: First, the finite duration, $\tau$, of collisions in the physical system leads to a finite rotation of the inter-particle unit vector. Consequently, the directions of the final velocities differ for MD (based on forces) and eMD (based on the coefficient of restitution). As indicated in the figure, the deviation may be large. Only for the case of a central collision, the directions of the final velocities agree for MD and eMD.

Second, the position of the particles as a function of time may be different for MD and eMD. This applies to both, central and off-central collisions. In Fig.~\ref{fig:illustrateCRule} we indicate the dynamical properties by plotting dots (of the respective color) on top of the trajectories (lines) at equidistant intervals of time. 

We wish to mention that the chosen parameters for the plot in \figRef{fig:illustrateCRule} correspond to very soft particles in order to visualize the differences between MD and eMD. A careful analysis \cite{mueller2011,negCONR} shows that the differences may be large also for more realistic material and system properties. The fundamental problems detailed above are \emph{always} present when collisions of physical particles are modeled by eMD assuming instantaneous collisions.

\subsection{Improved event-driven Molecular Dynamics}
\label{sec:impr-event-driv}

\subsubsection{Classification of events}
\label{sec:class-events}
The propagation rule, Eq.~\eqref{eq:cRuleHS}, is used in traditional eMD and relies on the coefficient of restitution, $\varepsilon_n$, and instantaneous collisions. It shall now be replaced by the propagation rule Eq.~\eqref{eq:cRule} using the matrix of restitution, $\tilde{\varepsilon}$, which takes the finite duration of collisions into account. We propose an improved eMD algorithm which does not suffer from the problems described above, caused by the assumption of instantaneous collisions.

%\patric{Der obige Satz erschien mir recht lang. Zudem taucht unsere Kollisionsroutine nie in Form einer geschlossenen Gleichung auf. Wir sollten deshalb eqref durch einen Verweis auf das entsprechende Kapitel ersetzen. Hier mein Vorschlag:}
%Traditional eMD relies on the coefficient of restitution, $\varepsilon_n$, as well as instantaneous collisions. The corresponding propagation rule is given in Eq. \eqref{eq:cRuleHS}. This collsion rule shall now be replaced by the propagation rule described in \secRef{sec:cRule} which uses the matrix of restitution, $\tilde{\varepsilon}$, and takes into account the finite duration of collisions. We propose an improved eMD algorithm which does not suffer from the above problems caused by the assumption of instantaneous collisions.

In the improved eMD scheme (eMD$^*$), each collision is represented by 3 instantaneous events. These events may be of type $\text{E}^\varphi$ or $\text{E}^v$:
 Assume two particles $(i,j)$ collide at time $t^*$. This collision is represented by
\begin{enumerate}
\item [a)] an event of type $\text{E}^\varphi$ at time $t^*$ where the positions of the particles are set due to the rotation of the inter-particle unit vector. The velocities are set to the center-of-mass velocity of the colliders, 
% \patric{Ich bin mir nicht sicher: Punkte am Ende der items?}
\item [b)] an event of type $\text{E}^v$ at time $t^*+\tau$ where particle $i$ adopts its post-collisional velocity, and
\item [c)] an event of type $\text{E}^v$ at time $t^*+\tau$ where particle $j$ adopts its post-collisional velocity.
\end{enumerate}

\subsubsection{Events of type $\text{E}^\varphi$}
\label{sec:events-tyoe}

An event of type $\text{E}^\varphi$ occurs at the moment $t^*$ when a pair of particles $(i,j)$ gets in contact, similar to the events in traditional eMD. The following sub-tasks are performed:
\begin{enumerate}
\item Compute the scaled parameters  $(\ln\tilde{l},\, \ln c_\varphi,\, \ln c_\text{dis})$ from the physical material parameters, the particle radii, the impact geometry and the velocities of the particles.
\item Compute the components $\left\{\varepsilon_\varphi,\varepsilon_t,\varepsilon_r,\varepsilon_{\dot{r}}\right\}$ of the matrix of restitution, \eqRef{eq:colMat}.  This may be done in a convenient and efficient way using lookup tables based on the transformations described in \secRef{sec:lookup}.
%\item Apply the collision rule detailed in \secRef{sec:cRule} using the matrix of restitution, $\tilde{\varepsilon}$, and the inverse scaling to obtain the rotation of the inter-center unit vector $\varphi=\varepsilon_\varphi \varPhi$, the postcollisional velocities, $\vec{v}_i^{\,\prime}$ and $\vec{v}_j^{\,\prime}$, and the duration of the collision, $\tau$, in physical coordinates.

%\patric{Die collisions routine in \secRef{sec:cRule} erledigt schon alles. Mein Vorschlag:

\item \label{item:rot} Apply the collision rule detailed in \secRef{sec:cRule} with $\varepsilon_t=0$ to rotate the particles around their center of mass by the angle $\varphi=\varepsilon_\varphi \varPhi$ and to obtain the postcollisional velocities, $\vec{v}_i^{\,\prime}$ and $\vec{v}_j^{\,\prime}$.
% \item \label{item:rot} Compute the positions of the particles by rotating them around their center of mass by the angle $\varphi$.
\item Set the velocities of both particles to the center of mass velocity $\vec{v}_i=\vec{v}_j=\dot{\vec{R}}^\text{0}$.
\item \label{item:storev} Store the computed post-collisional velocity $\vec{v}_i^{\,\prime}$ in a local variable of particle $i$ and, respectively, $\vec{v}_j^{\,\prime}$ in a local variable of particle $j$.
\item Mark both particles as {\em collision not yet accomplished} by setting a local flag.
\item Schedule two more events of type $\text{E}^v$ in the global event list, both occurring at time $t^*+\tau$; ($\tau=\varepsilon_t T$):
  \begin{enumerate}
  \item the velocity of particle $i$ will be updated.
  \item the velocity of particle $j$ will be updated.
 \end{enumerate}
\end{enumerate}

\subsubsection{Events of type $\text{E}^v$}
If a particle suffers an event of type $\text{E}^\varphi$ at time $t^*$, it suffers an event of type $\text{E}^v$ at time $t^*+\tau$, where $\tau$ is the duration of the collision which was computed when the event of type $\text{E}^\varphi$ was handled. In difference to events of type $\text{E}^\varphi$ describing two-particle interactions, events of type $\text{E}^v$ concern only one particle. The following sub-tasks are performed when an event of type $\text{E}^v$ occurs:
\begin{enumerate}
\item Check whether the flag {\em collision accomplished} is set in the concerned particle $i$. If this is the case, do nothing. Otherwise continue with item 2.
\item Set the velocity of the concerned particle $i$ to the value which was previously computed and stored in a local variable of particle $i$, see item \ref{item:storev} in Sec. \ref{sec:events-tyoe}.
\item Set the flag {\em collision accomplished} in particle $i$.
\end{enumerate}

\subsubsection{Schedule of events}
\label{sec:schedule-events}
The eMD$^*$ algorithm is similar to the eMD algorithm in the sense that the computation proceeds from one event to the next. The particle velocities are only changed due to these instantaneous events (except for the trivial acceleration resulting from homogeneous and constant external fields which does not influence the collision sequence). In eMD$^*$, the events of type $\text{E}^\varphi$ correspond to the events in eMD.

Again, we discuss only the principle of the algorithm, not the technicalities of its implementation. We assume there is a global list which contains the sequence of scheduled events of type $\text{E}^v$. Initially, the list is empty.

The  eMD$^*$ algorithm then works as follows:

\begin{enumerate}
\item find the next colliding pair $(i,j)$ of particles in the system and their collision time $t^*$ (begin of the collision)
\item if $t^*$ is smaller than the first (next in time) entry in the collision list, 
  propagate all particles $k$ to this time,
  \begin{equation}
    \label{eq:propStar}
    \vec{r}_k := \vec{r}_k+\vec{v}_k(t^*-t)\,,
 \end{equation}
handle the collision as an event of type $\text{E}^\varphi$ and proceed with step 1.
\item propagate all particles $k$ to the time $t^\dag$ of the next scheduled event of type $\text{E}^v$
  \begin{equation}
    \label{eq:7}
    \vec{r}_k := \vec{r}_k+\vec{v}_k(t^\dag-t)\,,
 \end{equation}
handle this event and remove the entry from the list. If there is more than one event scheduled for the same time, chose any of them. Proceed with step 1.
\end{enumerate}

An exemplary application of the algorithm is shown in Fig.~\ref{fig:illustrateCRule} (bottom). The black lines again denote the trajectories due to Newton's equation (same as upper panel). The green lines display the trajectories as obtained by the eMD$^*$ algorithm. At time {\em 0} when the particles get in contact, an event of type $\text{E}^\varphi$ is performed. This event rotates the inter-center unit vector around their center of mass, and, thus, relocates the particles instantaneously to new positions (time {\em 0} is shown twice). From there on, the particles move at the velocity of the center of mass. At time {\em 7} two events of type $\text{E}^v$ occur where both particles adopt their final post-collisional velocities. From this time on the trajectories due to eMD$^*$ and MD (Newton's equations) agree perfectly. In contrast, as indicated by the upper panel of Fig.~\ref{fig:illustrateCRule} the results of eMD (red lines) and MD differ significantly.

\subsubsection{Exceptions}
\label{sec:exceptions}

For the description of the algorithm, we silently assumed that the operations due to events of types $\text{E}^\varphi$ and $\text{E}^v$ are permitted. This is, however, not always the case but we have to deal with two possible exceptions:
\begin{enumerate}
\item[a)] The rotation step (event of type $\text{E}^\varphi$, item \ref{item:rot} of \ref{sec:events-tyoe}) may not be executed as it would lead to overlap with other particles. \label{item:rotxcept}
 \item[b)] In the time interval between an event of type $\text{E}^\varphi$ and the associated events of type $\text{E}^v$ the colliding particles $(i,j)$ move at the center of mass velocity, which is an unphysical but very short-lived transient state. In this time interval, one of the particles $(i,j)$ (or both) may collide with another particle $k$.
\end{enumerate}

{\em Case a):} We are of the opinion that event-driven MD is restricted to the domain of dilute systems. To apply eMD, we have to assure that in the corresponding physical (force-based) system, the frequency of three-particle interactions is negligible as compared with the frequency of two-particle interactions (see \cite{GranularGases} for a detailed discussion of this problem). For realistic and relevant material and system parameters, the rotation angle of the inter-center unit vector, $\varphi$, is rather small. Consequently, only minimal extra space is needed to perform this rotation. The probability that this small rotation would lead to overlap with other particles is, hence, small as well. 

In case that such an exception occurs, we fall back to the traditional eMD scheme for this particular collision: Only the velocities of the particles are changed according to the collision rule, Eq. \eqref{eq:cRuleHS}, using the coefficient of restitution $\varepsilon_n=-\varepsilon_{\dot{r}}$ but not the full matrix of restitution.

{\em Case b):} Assume the collision $(i,j)$ at time $t=0$ requires for both particles $i$ and $j$ an event of type $\text{E}^v$ at time $\tau$. Assume further that another particle $k$ collides with $i$ at time $t^k<\tau$. In this case we exceptionally perform the event of type $\text{E}^v$ of particle $i$ at time $t^k$, just before the event of type $E^\varphi$ of the pair $(i,k)$. 

That is, only the instant in time when the events of type $\text{E}^v$ are executed is modified. The post-collisional velocities are not affected by the exception handling and, hence, neither conservation of momentum nor conservation of energy are violated by this type of exception. But, as a consequence of the exception, the events of type $\text{E}^v$, \emph{both} scheduled at time $\tau$ originally, are no longer executed simultaneously due to the interference of a third particle. This, in turn, violates conservation of angular momentum by a tiny amount. However, it may be shown, that \emph{any} application of periodic boundary conditions leads to much stronger violations of angular momentum.

The tiny violation of the conservation of angular momentum may actually be avoided: The events $\text{E}^v$ of \emph{both} particles $i$ and $j$, originally scheduled at time $\tau$ are executed at the earlier time $t^k$ when either particle $i$ or $j$ interferes with a third particle $k$. However, this requires post-collisional communication between the particles $i$ and $j$ being an algorithmic complication which may cause significant loss of computational performance. Depending on case specific demands, one has to decide between absolute accuracy and maximal algorithmic efficiency. But anyway, for dilute systems, being the scope of the eMD$^*$ algorithm, eMD$^*$ including post-collisional communication is still by orders of magnitude more efficient than force-based MD.

Furthermore, the exceptions of type b) implicate the question of how to deal with collisions being interfered by more than a third particle. For a gas considered here, the mean free time is much larger than the time lag between the events of type $E^\varphi$ and $\text{E}^v$, corresponding to the duration of a collision. Therefore, if the frequency of an exception is small ($\sim0.1\%$, see \secRef{sec:confRegion}), the probability of a four-particle interaction is even much smaller ($\sim10^{-6}$). Hence, these cases may safely be neglected.

%The algorithm correctly reassembles the postcollisional soft sphere trajectories. At least for dilute systems it allows for \emph{event driven} simulation of \emph{soft} spheres. It extends classical eMD for granular systems based on the hard sphere model to soft spheres:

\subsubsection{Confidence Regions of the eMD$^*$ algorithm}
\label{sec:confRegion}
Aim of the eMD$^*$ algorithm is to simulate soft spheres while maintaining the advantages of event-driven modeling which, in its traditional form, relies on hard sphere interaction. Of course, this goal may only be achieved if the unavoidable exceptions detailed in~\ref{sec:exceptions} are rare and hence negligible events. In this section we, therefore, assess the range of validity of the eMD$^*$ algorithm by providing statistics on the exception frequency. To this, we simulate a granular gas of $N=10\:000$ elastic particles (interaction-force \eqRef{eq:fNViscel}, $A=0~\text{s}$, $\rho_m=1140~\text{kg}/\text{m}^3$, $R=0.1~\text{m}$, $\nu=0.4$). As simulation setup we choose a periodic box of Volume $V_\text{sim}$, in which the particles are initially located on a crystal lattice, from which they are then released to move freely with random velocities distributed in a way that the resulting thermal velocity is about $2~\text{m}/\text{s}$.

\begin{figure}
 \includegraphics[width=0.9\columnwidth,clip]{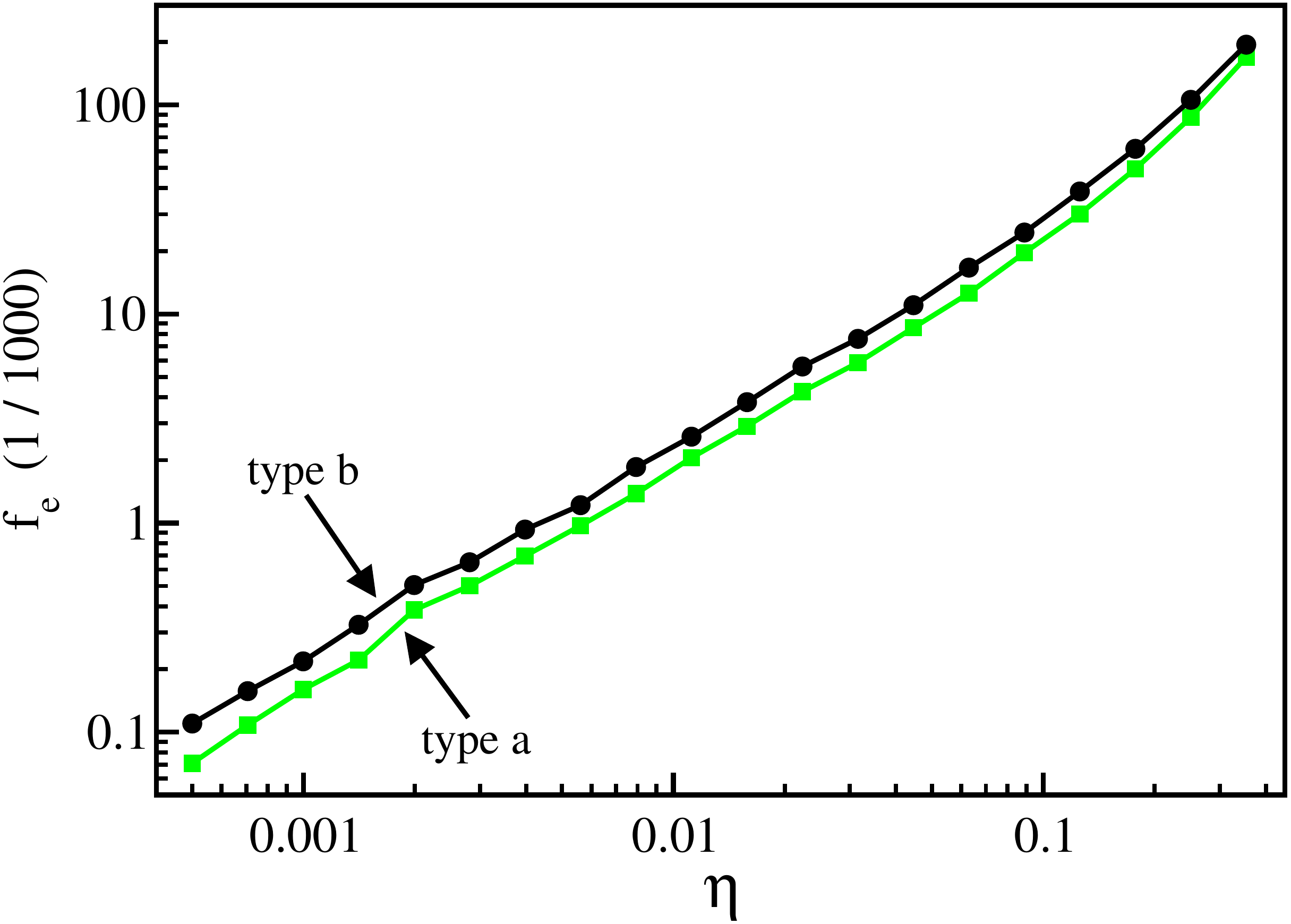}\\
\includegraphics[width=0.9\columnwidth,clip]{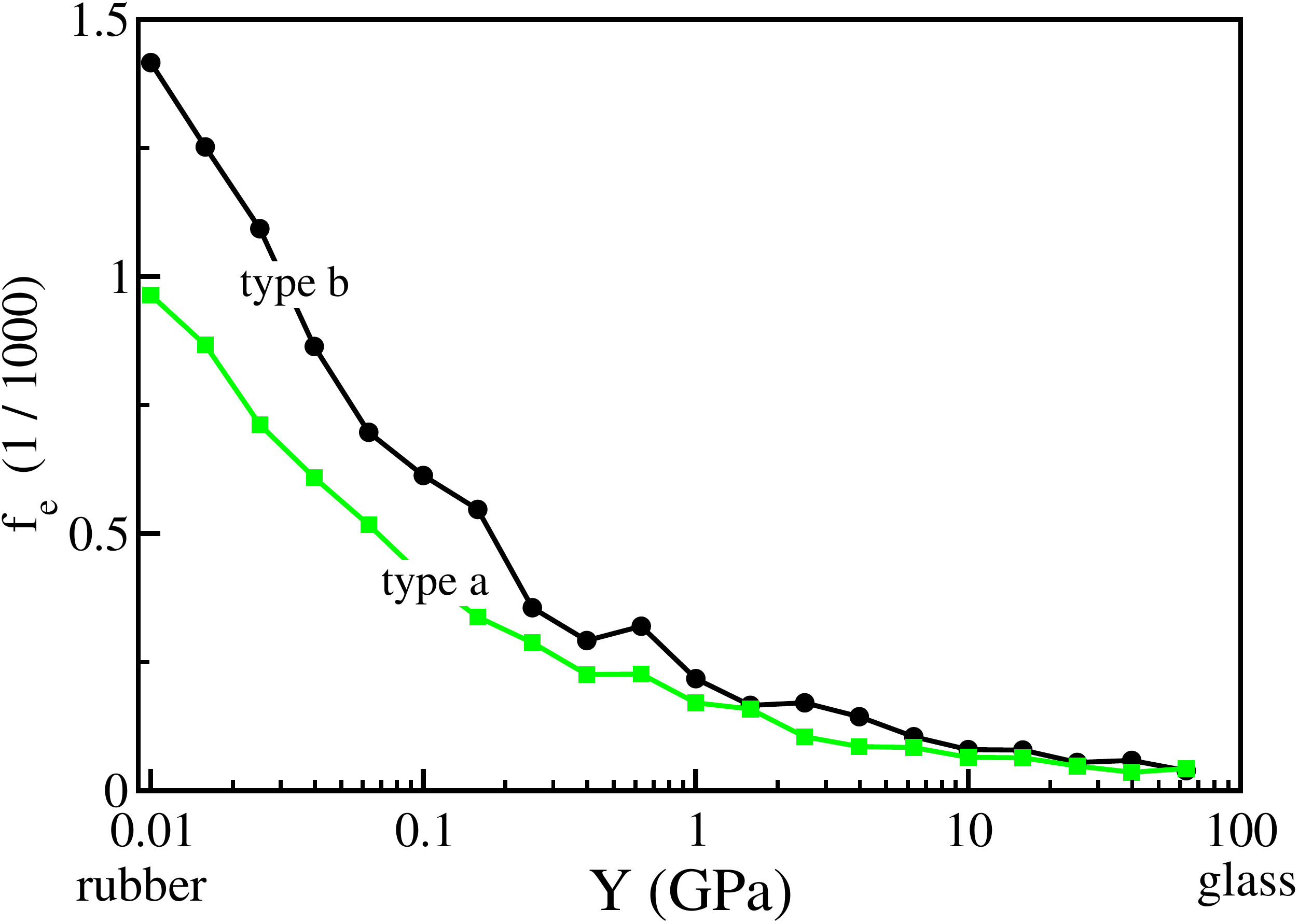}
\caption{(color online) Exception frequency $f_e$ in $\nicefrac{1}{1000}$ for a free granular gas of elastic particles as a function of a) the packing fraction $\eta$ at $Y=1~\text{GPa}$ (upper panel) and b)
the particles Young's modulus $Y$ at $\eta=0.001$ (lower panel).
See text for details on the setup and the parameters.}
 \label{fig:exceptions}
\end{figure}

Obviously, the frequency of both exceptions (type a, rotation step impossible, and b, three particle contact, see enumeration in~\ref{sec:exceptions} ) is mainly governed by the packing fraction
\begin{equation}
\eta\equiv\frac{N4 R^3\pi}{3V_\text{sim}}
\end{equation}
and the Young's modulus of the particle material, which, in turn, influences on the contact duration and the rotation angle $\varphi=\varepsilon_\varphi\Phi$, respectively. During the simulation we record the number of exceptions of type a) and b) for various packing fractions and Young's modulus ranging from very soft materials like e.g. rubber to hard materials like e.g. glass. The result is shown \figRef{fig:exceptions}. First we see that in the limit of very hard spheres or very dilute systems the probability of both exception types vanish. Second, for system parameters typically used in the literature on granular gases and for common materials, exceptions of both types are rare events (about $0.1\%$).

Clearly, the percentage of collisions where the eMD$^*$ algorithm fails (and we fall back to the traditional collision rule) is small. The eMD$^*$ algorithm, hence, indeed improves trajectory accuracy for typical systems.

\section{Summary}

Basic concept of event-driven Molecular Dynamics is the assumption of perfectly hard spheres leading to instantaneous collisions, such that the particle positions do not change during the collision. This assumption allows to describe the dynamics of a granular system as a series of independent binary collisions.
Each of these collisions is modeled by a simple multiplication of the pre-collisional relative velocity in normal direction with the coefficient of restitution to obtain the post-collisional value and finally the post-collisional vectorial velocities. The only parameter characterizing the collision is the coefficient of restitution, containing all the physics of the particle interaction.
For central collisions, it can be derived by integrating Newton's equation of motion for an isolated pair of colliding particles (which may lead to a velocity dependent coefficient of restitution), e.g. \cite{schwager2007,schwager2008,Schwager:1998,ramirez1999}. The coefficient of restitution is then found from its definition, Eq.~\eqref{eq:epsNHSDef}. Hence, for central collisions, an event-driven description yields the correct post-collisional velocities if compared to the integration of Newton's equation of motion. However, even for central collisions, the temporal properties are not correctly reproduced since the finite duration of collisions is neglected within event-driven modeling. 

Clearly, the assumption of instantaneous collisions is an approximation: Physically, the trajectories are determined by Newton's equation of motion with appropriate forces and material parameters. Any instantaneous change of the velocities would correspond to diverging repulsive forces between the particles. Furthermore, the request for a loss of energy of colliding particles (expressed by the coefficient of restitution) is not consistent with the assumption of instantaneous collisions since otherwise a finite amount of energy must be dissipated in vanishing time. That is, the hard-sphere model may be inappropriate for the description of dissipative systems.

The finite duration of physical collisions leads {\em always} to a finite rotation of the inter-particle unit vector $\vec{e}_r$. Only for central collisions $\vec{e}_r$ remains unchanged. For the case of adhesive nano-particles \cite{saitoh2010} it was recently shown, that at very large impact rate the rotation of $\vec{e}_r$ may be large. This rotation, in turn, causes a large deviation between the trajectories as obtained in eMD (applying the collision rule,  \eqref{eq:cRuleHS}) and MD (integrating Newton's equation). This result was generalized to oblique collisions of particles interacting via {\em any} force law \cite{mueller2011,negCONR}.

Consequently, due to the hard-sphere assumption, eMD agrees with MD neither regarding the spatial nor the temporal properties of the trajectories. The deviations may be large \cite{mueller2011,negCONR}.

In the present paper we propose an alternative event-driven algorithm, eMD$^*$. The essence of the eMD$^*$ algorithm is an extended collision rule. In contrast to the one of classical eMD, it changes not only the particle velocities but also their positions. Pre- and post-collisional states of the system differ in more than just the normal component of the relative velocity. We arrange all changing quantities in a vector which completely describes the system state. Compared to classical eMD, where pre- and post-collisional normal component of the relative velocity are related by the coefficient of restitution, pre- and post-collisional state vectors are, consequently,  related by a matrix within eMD$^*$. We termed this matrix {\em matrix of restitution}. Together with the concept of the state vectors, it allows to maintain the mathematical form of the hard sphere collision rule applied within classical eMD. Similar to the coefficient of restitution in eMD, all physical properties of the collision are mapped to the matrix of restitution. The eMD$^*$ algorithm does not assume instantaneous collisions. If applicable, the post-collisional particle positions and velocities obtained by eMD$^*$ agree with those obtained by integrating Newton's equations.
Algorithmically, in eMD$^*$ each collision is represented by 3 events of two different types which together map the pre-collisional state to the post-collisional one. 

Centerpiece of the method is the setup of the {\em matrix of restitution} as a functional of the particle interaction force law. We apply the eMD$^*$ algorithm to two examples which are important for practical applications, the linear-dashpot force and the viscoelastic Hertz force. Both force laws are characterized by two material properties. The geometry of the particles and the vectorial pre-collisional velocities are further parameters describing the impact. For both examples we demonstrate that the collision can be fully described by a set of three parameters which allows to represent the elements of the matrix of restitution in the form of \emph{universal} lookup tables. Using these tables the eMD$^*$ algorithm turns into a very efficient simulation method. 

We applied the eMD$^*$ algorithm to the oblique collision of two spheres and obtain identical post-collisional velocities as compared with Newton's equations. The trajectories are identical as well, except for a short-lived transient state whose duration is of the order of the duration of the collision. This means, that for dilute systems, where the exceptions detailed in~\ref{sec:exceptions} are rare, negligible events, eMD$^*$ simulations are equivalent with MD simulations. In fact, as shown in \ref{sec:confRegion}, the frequency of (algorithmic) failure may be reduced to any desired number by reducing the system density, while the physical effects of finite interaction forces are preserved. That is, both methods simulate granular systems composed of \emph{soft} spheres and yield the same trajectories as functions of time. At the same time, as an event-driven algorithm, eMD$^*$ is much more efficient than force-based MD. So far, we only considered frictionless interactions. This, however, is not a principal restriction and extending our findings to rough, frictional spheres is subject of future investigation.

%Disregarding the stepped potential approach (e.g. \cite{vanZon2008}), so far, the simulation of soft spheres was restricted to force based MD. 

Besides standard eMD, also the Kinetic Theory of granular gases is based on the hard sphere model since the Boltzmann equation is applicable only for hard spheres. This raises the question how the deviations between the trajectories obtained by means of the coefficient of restitution and from Newton's equation, affect the results of Kinetic Theory like, e.g., transport coefficients, which is subject of current research. For granular gases it is known that the vectorial particle velocities are correlated due to the dissipative nature of the interactions which necessarily implies a violation of molecular chaos \cite{soto2001,brito1998,pagonabarraga2001,poeschel2002}. It may hence be expected that the improved trajectory accuracy achieved by the eMD$^*$ algorithm is not screened by Molecular Chaos and leaves its fingerprints also in measurable macroscopic quantities like e.g. the coefficient of (self-)diffusion.    
 
\begin{acknowledgments}
The authors gratefully acknowledge the support of the Cluster of Excellence 'Engineering of Advanced Materials' at the University of Erlangen-Nuremberg, which is funded by the German Research Foundation (DFG) within the framework of its 'Excellence Initiative'. 
\end{acknowledgments}

% \bibliography{references.bib}

%Merlin.mbs v4.21 2009-07-09.
%

\end{document}